\documentclass[conference]{IEEEtran}
\IEEEoverridecommandlockouts

\usepackage{cite}
\usepackage{amsmath,amssymb,amsfonts}
\usepackage{algorithmic}
\usepackage{graphicx}
\usepackage{textcomp}
\usepackage{xcolor}
\def\BibTeX{{\rm B\kern-.05em{\sc i\kern-.025em b}\kern-.08em
    T\kern-.1667em\lower.7ex\hbox{E}\kern-.125emX}}

\usepackage[utf8]{inputenc}
\usepackage{booktabs}
\usepackage{tabularx}
\usepackage{makecell}
\usepackage{ragged2e}

\usepackage[utf8]{inputenc}
\usepackage{tikz}
\usepackage[draft,bookmarks=false]{hyperref}

\usetikzlibrary{shapes.geometric,arrows.meta,positioning,fit,backgrounds,shadows.blur,calc}
\definecolor{LegalNavy}{HTML}{00264D}
\definecolor{TechTeal}{HTML}{006666}
\definecolor{BridgePurple}{HTML}{5D2E8C}
\definecolor{AssetGreen}{HTML}{145214}
\definecolor{ZoneGray}{HTML}{F8F9FA}
\definecolor{AlertRed}{HTML}{C0392B}

\begin{document}

\definecolor{lightred}{rgb}{0.9,0.2,0.2}
\newcommand{\change}[1]{{\color{lightred} #1}}

\setlength{\textfloatsep}{4pt plus 1.0pt minus 1.0pt}
\setlength{\floatsep}{4pt plus 1.0pt minus 1.0pt}
\setlength{\abovecaptionskip}{2pt}
\setlength{\belowcaptionskip}{2pt}

\setlength{\abovedisplayskip}{2pt}  
\setlength{\belowdisplayskip}{6pt}  
\setlength{\abovedisplayshortskip}{1pt}  
\setlength{\belowdisplayshortskip}{1pt} 

\title{\LARGE SoK of RWA Tokenization: A Systematization of \\ Concepts, Architectures, and Legal Interoperability
}


\author{\IEEEauthorblockN{Junliang Luo\IEEEauthorrefmark{1}\IEEEauthorrefmark{10},
Xihan Xiong\IEEEauthorrefmark{3}\IEEEauthorrefmark{10},
Zonglun Li\IEEEauthorrefmark{1},
Hong Kang\IEEEauthorrefmark{1},
Xue Liu\IEEEauthorrefmark{1}\IEEEauthorrefmark{4},
William J Knottenbelt\IEEEauthorrefmark{3},
Katrin Tinn\IEEEauthorrefmark{2}
}
\vspace{0.1cm}
\IEEEauthorblockA{
School of Computer Science, McGill University, Montréal, Québec, Canada\IEEEauthorrefmark{1}\\
Desautels Faculty of Management, McGill University, Montréal, Québec, Canada\IEEEauthorrefmark{2}\\
Department of Computing, Imperial College London, London, United Kingdom\IEEEauthorrefmark{3}\\
Mohamed bin Zayed University of Artificial Intelligence, Abu Dhabi, UAE\IEEEauthorrefmark{4}\\
\IEEEauthorrefmark{10}Contributed equally to this work.
junliang.luo@mail.mcgill.ca, xihan.xiong20@imperial.ac.uk
}
}


\maketitle

\begin{abstract}
The global financial architecture is undergoing a shift from intermediary centric-settlement to programmable infrastructure, to transmute trillions in static illiquid capital into active, high-velocity instruments.
We argue that Real World Asset (RWA) tokenization represents a conceptual evolution beyond mere digitization, converting passive ledger entries into programmable economic agents capable of autonomous settlement and algorithmic collateralization.
However, achieving such seamless capital efficiency necessitates resolving the fundamental friction between deterministic on-chain code and probabilistic off-chain reality, navigating the oracle problem and jurisdictional interoperability.
This systematization of knowledge presents a taxonomy for the RWA lifecycle and deconstructs the multi-layered architecture, spanning legal custody, technical standards, and cryptoeconomic valuation, required to enforce off-chain rights within on-chain environments.
We study systemic constraints such as latency and regulatory fragmentation through a comparative overview of sovereign debt, private credit, and real estate protocols, complemented by an empirical case study of on-chain U.S. Treasuries.
We synthesize these findings to propose a prognostic outlook, positing that while asset tokenization provides a transitional bridge, it is not necessarily the inevitable shift compared to the emergence of unified, programmable ledgers.
\end{abstract}


\section{Introduction}
The global financial architecture is currently navigating a structural phase transition, an infrastructural transition, i.e., a reorganization from intermediary-centric, batch-settled systems to programmable settled financial infrastructure.
For centuries, the foundational layer of the global economy comprising real estate, private credit, commodities, and intellectual property, has existed as what economist Hernando de Soto characterizes as \textit{dead capital} \cite{desoto2000mystery}. 
These assets, while possessing immense intrinsic value estimated in the hundreds of trillions \cite{bcg2022tokenization, wef2025tokenization}, remain legally encumbered, geographically isolated, and operationally inert \cite{imf2023defi}. 
They function as passive stores of value, disconnected from the velocity of transactional markets and incapable of interacting without significant intermediary friction and verification costs \cite{coase1937nature, north1990institutions, catalini2016economics}.
Real World Asset (RWA) tokenization is reductively, analyzed as a mere digitization of existing financial instruments, a modernization of record-keeping akin to the migration from paper certificates to electronic databases \cite{oecd2020fragmentation}. 
We argue that this perspective is insufficient to capture the phenomenon. 
RWA tokenization represents a potential ontological shift: the transition of assets from passive ledger entries into \textit{programmable}, active economic agents \cite{bis2023unified, harvey2021defi}. 
By embedding legal rights, compliance logic (e.g., ERC-1400\cite{erc1400whitepaper}, ERC-3643 \cite{erc3643whitepaper}), and cash flow distribution directly into self-executing smart contracts, tokenization transmutes static capital into active assets capable of autonomous settlement, atomic composition, and algorithmic collateralization \cite{fed2023payments, cong2021tokenomics, erc3643whitepaper}.
This transition signifies a potential liquidity singularity: a theoretical state where the historic dichotomy between \textit{currency} (liquid, medium of exchange) and \textit{asset} (illiquid, store of value) effectively dissolves \cite{citi2023money}. 
In this context, the concept of \emph{money} expands to encompass any tokenized representation of value that can be verified and liquidated on-chain. 
Through the hyper-collateralization of reality, a fraction of a commercial building in New York or a freight invoice in Singapore can theoretically serve as instant, trustless collateral for liquidity in a global DeFi protocol \cite{maker2022endgame, centrifuge2021whitepaper, goldfinch2021whitepaper}. 
This fundamentally alters the physics of capital efficiency, allowing value to flow with the viscosity of information \cite{fink2024letter, mckinsey2023wealth}.
However, realizing this vision of a breathing global liquidity pool requires navigating a labyrinth of complex exogenous constraints. Unlike crypto-native assets (e.g., Bitcoin or Ethereum), which are governed solely by endogenous, deterministic code, RWAs must maintain a continuous, trust-minimized synchronization between the objective reality of the distributed ledger and the subjective, probabilistic reality of the physical world \cite{werbach2018trust}. 
This introduces the Oracle Problem \cite{caldarelli2020oracle} at an institutional scale (e.g., the inability of a blockchain to intrinsically verify if a physical warehouse has been destroyed or a legal contract voided), preserving that on-chain states accurately reflect off-chain legal truths regarding ownership, physical condition, and insolvency \cite{imf2023defi, goldfinch2021whitepaper}.
The integration of these time-varying assets into public ledgers creates friction with existing regulatory frameworks designed for static intermediaries. The challenge is not merely technical but jurisdictional, requiring a Law of the Code \cite{lessig1999code} that creates interoperability between the rigid enforcement of smart contracts and the flexible interpretation of judicial courts \cite{ buterin2022sbt}. 
Significant strides are being made in this domain through initiatives like the Monetary Authority of Singapore's Project Guardian, which tests the feasibility of asset interoperability across regulated entities \cite{mas2023guardian}.
We target a comprehensive systematization of the technical infrastructure, cryptoeconomic mechanisms, and legal frameworks underpinning the RWA paradigm. 
A taxonomy for RWAs is established to delineate their conceptual boundaries and map the complete lifecycle from origination to final settlement. 
We present a quantitative overview of the global market, examining the valuation frameworks and yield accrual mechanisms that govern on-chain price discovery. 
The multi-layered architecture required to bridge off-chain markets with on-chain ledgers is deconstructed by analyzing the critical interplay between legal custody, technical standards, and compliance enforcement. 
Systemic constraints inhibiting adoption, such as regulatory fragmentation and data latency, are scrutinized alongside a comparative analysis of incumbent protocols across sectors including sovereign debt, private credit, and real estate. 
We complement this theoretical study with an empirical analysis of on-chain tokenized U.S. Treasuries use case and conclude with a prognostic outlook suggesting that tokenization may not necessarily be the inevitable shift. 

\section{Definition and Conceptual Positioning of RWA}

\subsection{Definition and Core Essence}
Real World Assets (RWAs) are formally defined as the cryptographic embed of economic rights, spanning tangible property and intangible claims, within a distributed ledger base. 
Unlike native digital assets, which derive existence solely from the blockchain's consensus, RWAs operate as a dual-state system, requiring a continuous, legally binding synchronization between an off-chain registry (the legal truth) and an on-chain smart contract (the computational truth) \cite{he2017fintech}. 
This architecture resolves into the following functional primitives:

\subsubsection{Asset State Synchronization}
The foundational primitive is the establishment of a Ricardian Contract \cite{grigg2004ricardian}, where the digital token acts not only as a receipt, but as a legally operative executable of the underlying asset agreement. 
This synchronization requires an on-chain–off-chain mapping where every change in the physical asset status (e.g., depreciation, insurance expiry) is reflected in the token's metadata state without human intervention. 
This automated coupling reduces the verification cost that typically plagues the securitization of non-standard assets, converting heterogeneous physical risks into homogeneous digital risk \cite{catalini2016economics}.

\subsubsection{Programmability of Values}
The asset transitions from a passive database entry to an active computational object. 
The token inherits the general-purpose execution logic of the hosting chain, i.e., it can encode and automatically execute predefined contractual conditions, enabling it to self-execute financial lifecycle events such as automatic dividend distributions, principal amortization, or redemption at maturity, directly through its bytecode \cite{harvey2021defi}. 
The programmability of value enables RWAs to interact directly with decentralized finance (DeFi) protocols for lending, collateralization, and liquidity provision, meaning that the tokenized asset can autonomously participate in borrowing and lending markets, be posted as on-chain collateral, and supply liquidity to market-making mechanisms, with contractual cash flows, ownership constraints, and settlement rules enforced natively at the protocol level rather than through off-chain intermediaries \cite{garratt2023continuum}.

\subsubsection{Settlement and Finality}
Another primitive is the replacement of sequential clearing cycles with atomic delivery-versus-payment (DvP) \cite{pinna2016dlt}. 
In traditional infrastructure, the transfer of an asset and the movement of funds occur on separate, asynchronous rails, introducing counterparty and settlement risk \cite{malinova2023tokenomics}. 
In tokenised real-world asset systems, execution and settlement can be collapsed into a single atomic transaction in which the asset transfer is cryptographically contingent on the simultaneous success of the payment leg, thereby mitigating principal risk \cite{pinna2016dlt}. 
Regulatory frameworks such as the EU DLT Pilot Regime \cite{esma2022dlt} establish a controlled environment for DLT-based trading and settlement infrastructures that could support such settlement models for tokenised financial instruments while preserving market integrity.

\subsection{Conceptual Boundaries and Differentiation}
Taxonomy demarcation is required to distinguish RWAs from adjacent digital asset classes. While tokenization describes the broad technological process of imprinting rights onto a blockchain, RWA specifically refers to the asset class of real world asset itself. 
This distinction resolves into three primary boundary definitions:

\subsubsection{Instrument vs. Numeraire (RWAs vs. Stablecoins)}
While both asset classes utilize similar technical standards (ERC-20 for the most stablecoins and RWA tokens), they serve orthogonal economic functions. 
Stablecoins are designed as numeraires to minimize volatility and function as a medium of exchange \cite{gorton2023taming}, the M0 money, i.e., the monetary base, consisting of physical currency or central bank reserves held by commercial banks.
In contrast, RWAs function as investment instruments whose on-chain representations encode exposure to yield, capital appreciation, or utility rights, with risk profiles ranging from cash-equivalent to risk-bearing assets.
These rights introduce market risk and duration exposure distinct from the par-value preservation mechanisms of stablecoins, separating the function of value from unit of account \cite{bullmann2019search}.

\subsubsection{Underlying Exposure vs. Regulatory Envelope (RWA vs. STO)}
Security Token Offering (STO) refers to the market issuance mechanism, i.e., the regulatory envelope used to distribute tokens in compliance with securities law (e.g., Reg D, Reg S) \cite{momtaz2022sto}. 
RWA, defines the ontological status of the underlying economic exposure. 
An RWA is the asset, e.g., the real estate equity, treasuries, power supply credits) \cite{laschinger2024liquidity, peng2025current, kong2025real}.
Thus, all STOs are mechanisms for token distribution, while the RWA concept extends beyond issuance to include lifecycle management and secondary market collateralization \cite{hacker2018crypto}.

\subsubsection{Digital vs. Analog Securitization (RWA vs. ABS/REIT)}
Functionally, RWAs replicate the logic of traditional Asset-Backed Securities (ABS), i.e., financial instruments collateralized by a pool of income-generating assets such as loans or leases \cite{fabozzi2021handbook}, or Real Estate Investment Trusts (REITs), i.e., corporate entities that own, operate, or finance income-producing real estate \cite{block2011investing}). 
This replication is achieved by pooling cash flows into a bankruptcy-remote Special Purpose Vehicle (SPV), i.e., a subsidiary legal entity created strictly to isolate financial risk and secure the underlying assets from the issuer's insolvency \cite{gorton2007spv}. 
However, RWA differentiates itself through operational transparency and composability. 
Unlike opaque ABS tranches, RWA tranches offer real-time, on-chain verification of the underlying loan performance. This transparency creates a Glass-Box Securitization model where credit risk is auditable by any participant in real-time \cite{makarov2022crypto}.

\subsection{Theoretical and Practical Relevance}
The relevance of integration of off-chain assets into on-chain environments is observed across three functional aspects:
\subsubsection{Bridging Function}
RWA tokenization serves as the connective intermediate between the high-compliance architecture of Traditional Finance (TradFi) and the permissionless liquidity rails of Decentralized Finance (DeFi). 
This creates an interoperability layer where institutional capital (e.g., pension funds) can access yield from decentralized lending protocols without violating regulatory mandates, provided the assets are wrapped in compliant standards \cite{avgouleas2019architecture} such as ERC-3643, the token standard for RWA tokenization \cite{erc3643org}. 
This bridge functions bi-directionally: it allows DeFi protocols to stabilize their treasuries with non-correlated real-world collateral (e.g., U.S. Treasury) while offering TradFi institutions access to always-on global settlement infrastructure \cite{iosco2023defi}.

\subsubsection{Financial Innovation: Democratization via Fractionalization}
From a microeconomic perspective, RWAs lower the barriers to entry for high-value asset classes through hyper-fractionalization. 
By dividing indivisible assets (e.g., a \$50 million commercial building) into fungible tokenized shares, the technology reduces the minimum ticket size for investment, thereby deepening market liquidity and democratizing access for retail participants \cite{chen2020blockchain}. 
Furthermore, the innovation enables continuous markets, moving asset trading from limited banking hours to an always-on trading, which theoretically reduces the liquidity premium associated with private market assets by allowing for instant exit and price discovery \cite{malinova2023tokenomics}.

\subsubsection{Systemic Significance of Unified Ledgers}
At a macro-prudential level, RWA tokenization establishes the foundation for the \textit{Finternet} or a \textit{Unified Ledger}: a shared digital infrastructure where multiple financial assets and currencies reside on the same programmable platform \cite{carstens2024finternet}. 
This creates a closed loop from asset origination to final settlement. 
Unlike the current fragmented system where securities, cash, and derivatives reside on separate proprietary databases requiring costly messaging synchronization (e.g., SWIFT), Unified ledgers allow for the atomic exchange of any asset against any other, effectively turning the entire balance sheet of the global economy into a liquid, programmable state \cite{bis2023unified}.
%

\section{RWA Taxonomy}
\begin{figure}[htb]
    \centering
    \includegraphics[width=\linewidth]{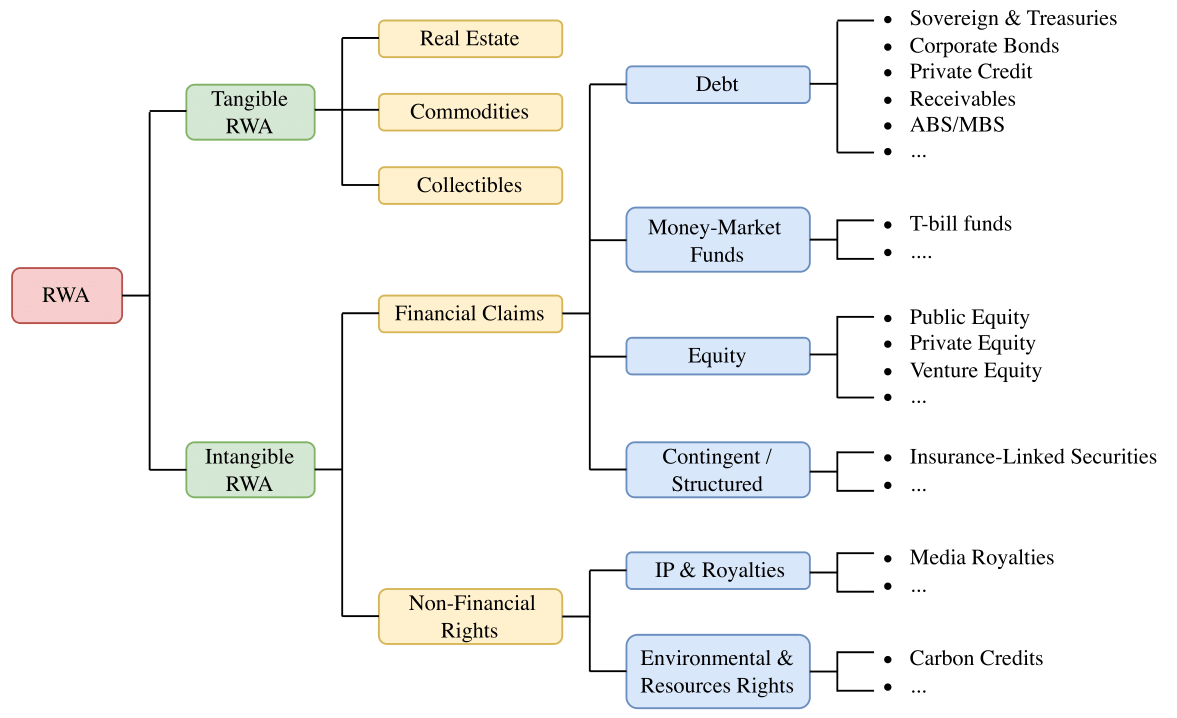}
    \caption{RWA: taxonomy by reference asset type.}
    \label{fig:taxonomy}
\end{figure}

In this section, we propose a taxonomy for RWAs (see Figure~\ref{fig:taxonomy}), focusing on the nature of the reference asset and on the legal rights attached to it. 
The first and most basic division separates tangible from intangible assets. This is a binary distinction that depends only on whether the asset has a physical form. Within the intangible group, the taxonomy introduces a second distinction. 
Some assets grant explicit contractual claims to future cash flows, whereas others confer rights of use, licensing, or entitlement whose value is not tied to fixed payments. The two groups are labeled ``financial claims'' and ``non-financial rights''. 

\subsection{Tangible RWA}

Tangible RWAs include assets whose economic value is closely tied to physical possession or control. The most visible category is real estate. Here tokenization usually represents fractional interests in income-producing property, either residential or commercial. Platforms such as \textit{RealT}~\cite{realt} and \textit{Lofty}~\cite{lofty} use special-purpose vehicles to hold U.S. rental property and distribute rental income on-chain.


Commodities form a second tangible category in RWA markets, with activity concentrated primarily in precious metals due to storage, auditability, and custody constraints. Instruments such as \textit{PAX Gold}~\cite{paxos-paxg} and \textit{Tether Gold}~\cite{tether-xaut} are representative examples, backed by vaulted bullion and offering redemption mechanisms through professional custodial arrangements operating under established regulatory regimes.

A smaller but conceptually important segment concerns collectibles (e.g., art pieces and luxury items), where tokenization facilitates fractional exposure to high-value assets. Initiatives such as \textit{Maecenas}~\cite{maecenas} illustrate a basic tokenization model that enables fractional investment in otherwise illiquid assets, although secondary-market liquidity remains limited, and valuations rely heavily on specialized appraisal processes.

\subsection{Intangible RWA}

\subsubsection{Financial Claims}
Intangible RWAs are more diverse. The largest group consists of financial claims, where cash flows are defined ex ante by contract. Debt instruments dominate this segment. Tokenized sovereign exposure has grown rapidly through structures that hold U.S. Treasuries or government money-market portfolios. Examples include products issued through \textit{Franklin Templeton}’s~\cite{franklin-benji} on-chain government fund and \textit{BlackRock}’s~\cite{blackrock-buidl} institutional liquidity fund. Corporate debt instruments have been issued in digital form on regulated platforms, such as \textit{Société Générale Forge}~\cite{sgf-greenbond}, which provide on-chain issuance and settlement infrastructure under existing regulatory frameworks. Private credit markets have also migrated on-chain, with protocols such as \textit{Maple}, \textit{Centrifuge}~\cite{centrifuge}, and \textit{Goldfinch}~\cite{goldfinch} extending credit to small firms, real-estate borrowers, and other off-chain entities. Receivable financing follows a similar pattern and is often embedded in structured pools. In parallel, some collateral systems integrate tokenized claims on asset-backed securities (ABS), illustrating how traditional securitization can connect to decentralized financial infrastructure.

Money-market funds (MMFs) are collective investment vehicles whose underlying assets consist primarily of short-term, high-quality debt instruments, such as Treasury bills, repurchase agreements, and commercial paper. Although their portfolios are dominated by sovereign and high-grade debt, MMFs are conceptually distinct from direct debt claims, as investors hold shares in a pooled vehicle rather than direct ownership of individual government bonds.
This model is visible in several high-profile implementations. \textit{Franklin Templeton}~\cite{franklin-benji}, through its Benji Technology Platform, has launched tokenized shares of regulated MMFs that invest primarily in short-term government and high-quality debt instruments. \textit{BlackRock}’s tokenized liquidity fund~\cite{blackrock-buidl}, known as BUIDL, similarly adopts this structure, offering stable-value tokens backed by Treasury bills and other safe, short-duration assets.

A third branch of financial claims covers Equity claims, which represent residual participation rather than fixed repayment. They appear in several forms: public equity issued on regulated digital securities venues, private equity structured through special-purpose entities, and venture or growth equity tied to early-stage firms. Platforms such as \textit{Securitize}~\cite{securitize} have issued multiple examples. Development in this area is relatively slow, as regulatory constraints and liquidity restrictions impose tighter limits on investor access and trading.

The final component within financial claims includes contingent and structured instruments. Their payoffs depend on specified events or on contractual structures rather than on simple interest schedules. Early initiatives in insurance-linked securities and catastrophe risk illustrate the potential for programmable ledgers to support complex risk transfer, although the market is still emerging and remains largely experimental.

\subsubsection{Non-Financial Rights}
Non-financial rights constitute the other main branch of intangible RWAs. In these cases, value arises from entitlement, licensing, or usage rather than from predefined interest payments. Intellectual-property-based structures are the most advanced. Several platforms issue tokens backed by royalties, where holders receive proportional revenue streams as works are used or streamed. Film and media rights follow similar models. Environmental and resource rights form another instance, most visibly in the tokenization of carbon credits by projects such as \textit{Toucan}~\cite{toucan}. 

\section{Lifecycle of RWA}

\begin{figure}[t!]
    \centering
    \includegraphics[width=\linewidth]{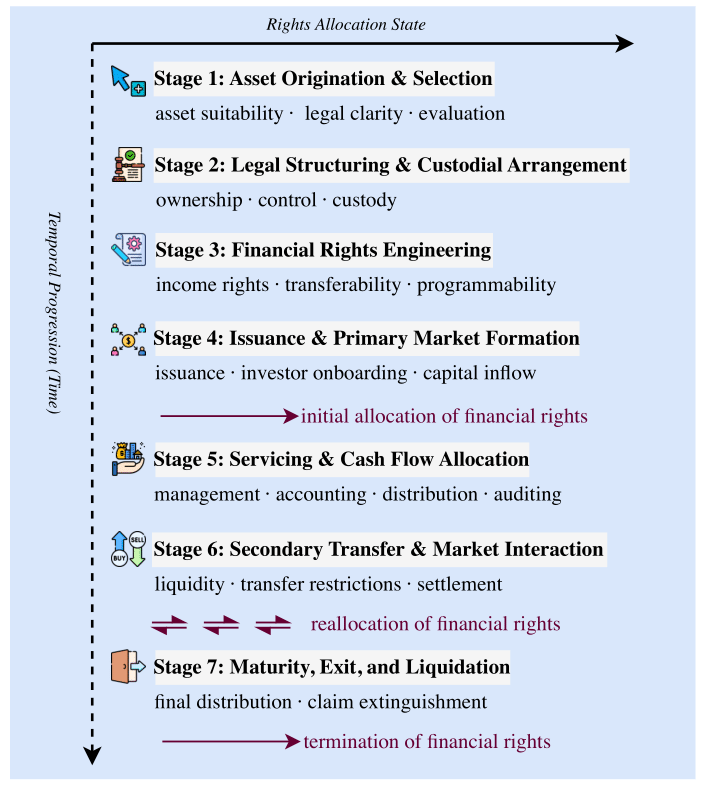}
    \caption{RWA: functional lifecycle framework.}
    \label{fig:lifecycle}
\end{figure}

This section presents a functional decomposition of the lifecycle of RWAs (see Figure~\ref{fig:lifecycle}), outlining the sequential stages through which real-world financial and physical assets are transformed into investable financial instruments on-chain.

\subsection{Stage 1: Asset Origination \& Selection}

Asset origination constitutes the initial stage of the RWA lifecycle, during which candidate assets are identified and assessed for their suitability for financialization. The primary objective of this stage is to determine whether an RWA possesses the economic, legal, and operational characteristics required for structured investment~\cite{mirdala2025tokenization,cfa_tokenization}. Typical evaluation criteria include the asset’s ability to generate predictable cash flows, the clarity and enforceability of ownership rights, valuation transparency, and legal compatibility. The origination process commonly involves asset owners, asset managers, and external due diligence providers such as accounting firms.

\subsection{Stage 2: Legal Structuring \& Custodial Arrangement}

Legal structuring represents a foundational stage in the RWA lifecycle, as it establishes the formal legal framework through which RWAs are isolated, owned, and administered~\cite{okuyama2025private}. The central objective is to transform a physical or contractual asset into a legally separable investment object, typically through the creation of a special purpose vehicle (SPV), trust, or fund structure. This process delineates ownership, control, and economic rights, while also addressing bankruptcy remoteness and investor protection~\cite{mirdala2025tokenization}. Custodial arrangements are determined at this stage, specifying which entities are responsible for the safekeeping and administration of the underlying assets and related cash flows. Key participants include legal counsel, custodians, compliance advisors, and regulators. 

\subsection{Stage 3: Financial Rights Engineering}
Financial rights engineering concerns the design and formalization of the economic rights associated with the reference asset. At this stage, legally defined claims are translated into standardized financial entitlements, such as rights to periodic income, principal repayment, residual value upon liquidation, or limited governance participation~\cite{cfa_tokenization}. The objective is to structure these rights in a manner that renders them intelligible, transferable, and investable within financial markets. This stage also involves determining priority structures, distribution schedules, and transferability constraints. In blockchain-based implementations, financial rights engineering may include the selection of token standards (see Table~\ref{tab:rwa_standards}) and smart contract architectures as they encode economic entitlements and compliance logic. Participants typically include asset managers, financial engineers, and accounting specialists. Importantly, this stage defines the nature of the financial product independent of its subsequent issuance or market distribution.

\subsection{Stage 4: Issuance \& Primary Market Formation}
The issuance stage marks the transition of engineered financial rights into the investment market through primary distribution. The primary objective is to allocate newly created financial claims to investors and to establish an initial valuation. This process involves the issuance of financial instruments and their sale to eligible participants under defined regulatory constraints. Investor onboarding and capital inflow into the asset-holding entity are central components of this stage. Typical participants include issuers, investors, distribution platforms, and compliance service providers responsible for know-your-customer (KYC) and eligibility verification~\cite{worldbank_tokenization}. The completion of primary issuance signifies the point at which the RWA formally enters the financial system.

\subsection{Stage 5: Servicing \& Cash Flow Allocation}
Servicing constitutes the operational core of the RWA lifecycle and encompasses the ongoing management of the reference asset and its associated cash flows. The principal objective of this stage is to ensure that economic value generated by the asset is accurately collected, accounted for, and distributed in accordance with predefined financial rights. Activities include asset operation, revenue collection, expense management, accounting reconciliation, and periodic income distribution to investors. In blockchain-enabled RWAs, cash flow allocation can be automated through smart contracts~\cite{worldbank_tokenization}, though reference asset performance remains anchored in off-chain processes. Key participants include asset servicers, property managers or obligors, accounting firms, auditors, and investors. This stage persists throughout the asset’s life and represents the primary source of realized investor returns.

\subsection{Stage 6: Secondary Transfer \& Market Interaction}
Secondary transfer refers to the post-issuance exchange of financial rights among investors~\cite{cfa_tokenization,worldbank_tokenization}. The objective of this stage is to provide liquidity and portfolio flexibility by enabling investors to reallocate exposure prior to asset maturity. Secondary market participation may occur through regulated exchanges or permissioned trading platforms, depending on regulatory and contractual constraints. While not all RWAs support active secondary markets, their existence enhances the financial characteristics of the instrument by reducing holding period rigidity. Participants in this stage include secondary investors, trading venues, custodians, and settlement systems. 

\subsection{Stage 7: Maturity, Exit, and Liquidation}
The final stage involves the termination of the financial arrangement through asset maturity, sale, or liquidation~\cite{worldbank_tokenization}. The objective is to convert remaining asset value into distributable proceeds and to extinguish outstanding financial claims. Exit mechanisms may be triggered by contractual maturity dates, strategic asset disposition, or predefined termination events. The liquidation process typically includes asset valuation, settlement of outstanding liabilities, and a proportional distribution of the residual value to investors, followed by the dissolution of the legal entity or the cancellation of financial instruments. Participants include asset managers, liquidators, custodians, and investors. A clearly defined exit stage is crucial for lifecycle completeness and for aligning investor expectations with legal and economic outcomes.

\section{Market Overview}

\subsection{Global Market Size and Growth}

\begin{figure}[htbp]
\centering
\resizebox{0.518\textwidth}{!}{%
    \includegraphics[width=\textwidth]{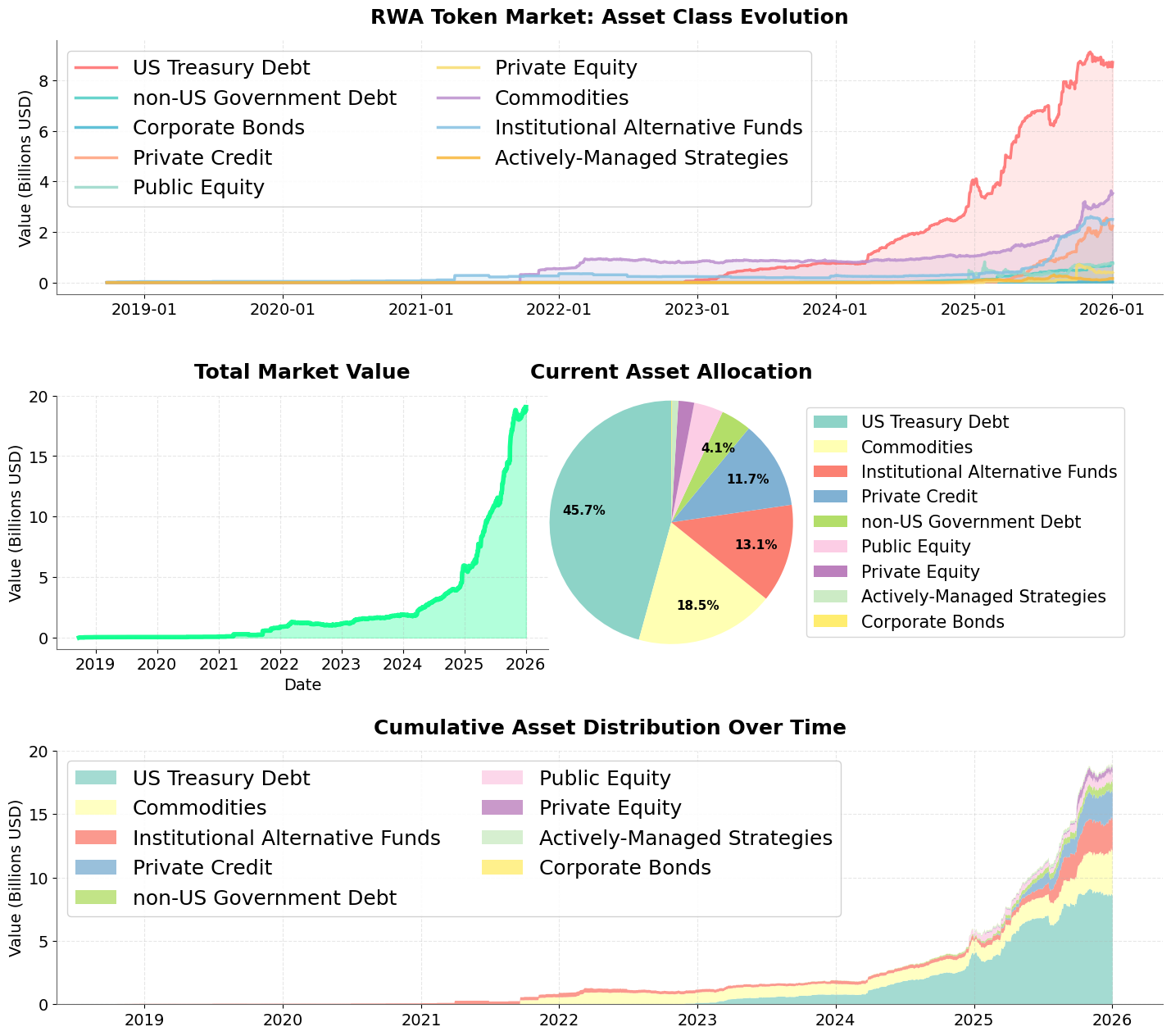}
}
\caption{Overview of the RWA token market from September 2018 to January 2026. The figure comprises four panels: the time-series evolution of nine major RWA asset classes, the aggregate total market capitalization over time, the current asset allocation across classes, and a stacked area chart showing the cumulative market composition and its structural changes over time. Together, the panels present a multi-dimensional view of market growth, asset dominance, and compositional shifts.}
\label{fig:rwa_market_analysis}
\end{figure}

As illustrated in Figure~\ref{fig:rwa_market_analysis}, the aggregate market capitalization has exhibited a non-linear growth trajectory, expanding from negligible volumes in 2020 to approximately \$19.38 billion by January 2026 \cite{rwa_xyz_2026}.
The compositional evolution of the market indicates a decisive shift toward yield-bearing debt instruments. While early tokenization efforts 2019-2020 were fragmented across niche commodities, the maturity phase 2023-2026 is characterized by the hegemony of fixed-income products. 
Specifically, private credit and tokenized treasuries have emerged as the dominant asset classes excluding commodities, collectively representing around 57\% of the total value locked \cite{rwa_xyz_2026}. 
Tokenized treasuries has solidified a \$8.7 billion market share, serving as the main collateral class. 
Conversely, tangible assets such as commodities remain competitive, accounting for approximately \$3.6 billion, about 18.5\% of the total RWA market at the latest observation point.

\subsection{Investor Profile and Yield Characteristics}
The observed market structure reflects a fragmented investor profile, with participation patterns varying across asset classes. 
Tokenized treasuries, exhibit strong institutional alignment, driven by compliance-heavy onboarding requirements, minimum investment thresholds, and suitability constraints that naturally favor professional investors. 
However, this institutional dominance is not uniform across the treasury segment itself. For example, products such as BUIDL (BlackRock USD Institutional Digital Liquidity Fund) \cite{BUIDL} are explicitly structured for institutional treasury management, whereas alternatives like Ondo’s USDY (U.S. Dollar Yield) \cite{ondo-usdy} adopt a more retail-accessible design \cite{luo2025transaction}, resulting in a bifurcated investor composition even within similar yield-bearing instruments.
Broadly, the RWA investor profile is characterized less by speculative positioning and more by yield stability and balance-sheet utility. 
Unlike crypto-native assets that attract high-beta trading behavior, RWA tokens are primarily utilized for capital preservation, predictable income generation, and collateral backing in on-chain financial activities. 
This distinction suggests that RWAs function as infrastructural financial primitives within decentralized ecosystems, serving treasury allocation and risk-managed liquidity deployment needs rather than acting as vehicles for short-term speculative returns.
%


\section{Valuation, Cash Flows, and Asset Price Dynamics in RWAs}

This section analyzes how RWAs are valued and how investor returns are realized over the lifecycle (see Table~\ref{tab:valuation-income-lifecycle}). We show that RWA valuation and returns are best understood through a lifecycle-based decomposition. Initial valuation governs the allocation of financial rights, servicing-phase income reflects realized cash flows, and reference asset price fluctuations primarily affect revaluation and terminal outcomes.


\begin{table}[htb]
\centering
\caption{Roles of valuation and income across lifecycle phases}
\renewcommand{\arraystretch}{0.85}
\label{tab:valuation-income-lifecycle}
\resizebox{\linewidth}{!}{
\begin{tabular}{lll}
\toprule
\textbf{Lifecycle Phase} & \textbf{Valuation Role} & \textbf{Income Role} \\
\midrule
Stage 4 (Issuance) & \makecell[l]{Initial pricing\\ (rights allocation)} & --- \\
Stage 5 (Servicing) & --- & \makecell[l]{Cash flow realization\\ and distribution} \\
Stage 6 (Secondary transfer) & \makecell[l]{Market revaluation\\ (rights reallocation)} & --- \\
Stage 7 (Maturity/Exit) & Terminal value realization & \makecell[l]{Final payout and\\ rights termination} \\
\bottomrule
\end{tabular}
}
\end{table}

\subsection{Conceptual Foundations: Valuation, Income, and Price}
Valuation is the present value of legally enforceable financial claims on a real-world asset, typically derived from expected future cash flows and a terminal value~\cite{modigliani1958cost}. Income denotes realized cash flows distributed to investors during the servicing phase, such as rental income, interest payments, or contractual repayments~\cite{jensen1986agency}. Market price, by contrast, reflects the transaction price at which financial claims are exchanged in primary or secondary markets and incorporates investor expectations, risk premia, and liquidity conditions~\cite{amihud1986asset}.

\subsection{Initial Valuation at Issuance}

Initial valuation occurs at the issuance stage (stage 4 in Figure~\ref{fig:lifecycle}), where financial rights over the reference asset are first allocated to investors. The object of valuation is not the token per se, but the standardized claim it represents~\cite{duffie2001term}. 
Across asset categories, the economic value of an RWA can be broadly expressed as the present value of expected servicing-phase cash flows plus terminal value, net of structural costs. The proposed model adopts a generalized discounted cash flow (DCF) framework, consistent with Damodaran’s proposition that the intrinsic value of any asset, regardless of its physical or digital form, is a function of its expected cash flows, their growth, and the associated riskiness~\cite{damodaran2012investment}.
Let $\mathcal{R}$ denote the financial rights represented by a token or fractional share. Initial valuation at issuance can be written as:

\begin{equation}
V_0(\mathcal{R})
= \mathbb{E}_0\Bigg[\sum_{t=1}^{T}\frac{CF_t(\mathcal{R})}{(1+r)^t}
+ \frac{RV_T(\mathcal{R})}{(1+r)^T}\Bigg] - C_0 ,
\end{equation}

$CF_t(\mathcal{R})$ denotes expected servicing-phase cash flows, $RV_T(\mathcal{R})$ the expected terminal value at maturity or liquidation, $r$ the discount rate reflecting risk, enforceability, and liquidity, and $C_0$ the structural costs associated with legal formation, custody, and ongoing administration. In this model, expected servicing-phase cash flows and the terminal value are projected from asset-specific characteristics and discounted at an appropriate risk-adjusted rate. Structural costs are deducted to determine the net investable value. The resulting valuation anchors the issuance price, which can be interpreted as the initial allocation price of financial rights. While issuance mechanisms may vary, the underlying economic logic remains consistent across RWA categories.

Across RWAs, heterogeneity in valuation arises primarily from differences in the relative importance of servicing-phase cash flows $CF_t$ and the terminal value $RV_T$. For analytical clarity, RWA structures can be broadly characterized by three stylized cash-flow patterns: (i) structures dominated by contractual cash flows (e.g., corporate bonds), (ii) structures relying on operating cash flows (e.g., real estate), and (iii) structures in which expected returns are primarily driven by terminal value realization (e.g., collectibles).

\subsection{Income Realization During the Servicing Phase}
Investor returns during the servicing phase (stage 5) are primarily realized through distributable cash flows rather than changes in valuation. Let gross inflows during period $t$ be denoted by $GI_t$. Distributable cash flow is computed as:

\begin{equation}
DCF_t
= GI_t - OE_t - SF_t - \Delta Res_t ,
\end{equation}

where $OE_t$ represents operating expenses, $SF_t$ servicing and management fees, and $\Delta Res_t$ changes in required reserves. For a structure with $N_t$ tokens outstanding, per-token distributions are typically given by:

\begin{equation}
d_t = \alpha \cdot \frac{DCF_t}{N_t},
\end{equation}

where $\alpha$ denotes the distribution ratio. 

Importantly, in most RWA designs, servicing income depends on realized operational or contractual performance rather than contemporaneous market valuations of the underlying asset. As long as the asset continues to operate and contractual obligations are met, periodic income remains largely insulated from short-term price fluctuations.

\subsection{Asset Price Fluctuations, Revaluation, and Termination}

Underlying asset price fluctuations play a limited role during the servicing phase but are central to revaluation and terminal outcomes. In secondary markets, trading prices reflect investors’ updated expectations regarding future distributions, discount rates, and terminal value~\cite{campbell1988dividend}. Abstractly, a secondary market price at time $t$ can be expressed as:

\begin{equation}
P_t
= \mathbb{E}_t\Bigg[\sum_{s=t+1}^{T}\frac{d_s}{(1+\rho_{t,s})^{s-t}}
+ \frac{\tilde{RV}_T}{(1+\rho_{t,T})^{T-t}}\Bigg],
\end{equation}

where $P_t$ is the secondary-market price of the RWA claim at time $t$; $d_s$ represents the per-unit distributable cash flow at future time $s$ generated during the servicing phase; $\rho_{t,s}$ denotes the market-implied required return between $t$ and $s$; and $\tilde{RV}_T$ is the expected terminal value. The conditional expectation is taken with respect to the information set available at time $t$.


At maturity or liquidation, reference asset prices directly determine terminal payouts. The realized sale or redemption value, net of exit costs and liabilities, defines the final distribution to investors and marks the termination of financial rights. Exceptions to this general pattern arise when cash flows are contractually linked to asset valuation or when valuation shocks trigger early liquidation or default, effectively shifting the structure from servicing to termination.

\subsection{Valuation Uncertainty and Structural Ambiguity}
While the valuation framework above provides a unified representation of RWA value formation, valuation outcomes are subject to structural sources of uncertainty. In practice, RWA valuation should often be interpreted as a range of plausible outcomes rather than a precise point estimate.

First, valuation uncertainty arises from the predictability of future cash flows. Contractual cash-flow RWAs allow relatively stable estimation, whereas operating cash-flow structures are exposed to variability in utilization, costs, and market conditions, increasing sensitivity to modeling assumptions and discount rates.
Second, in structures where expected returns are dominated by terminal value, valuation becomes highly sensitive to assumptions regarding exit timing, liquidity, and market conditions~\cite{titman1989valuing}. In such residual-value-dominant RWAs, discounted cash flow approaches offer limited anchoring power, as servicing-phase cash flows are weak or negligible.
Finally, RWAs face legal and enforceability risks inherent to their hybrid on-chain/off-chain structure~\cite{aloshyna2025tokenization,chen2024exploring}. The realizability of projected cash flows and terminal claims depends on the effectiveness of legal structuring, custody, and contractual enforcement, which may vary across jurisdictions.


\section{Stakeholder and Adoption Perspectives}
The adoption of tokenized RWA systems is not monolithic. The systems are shaped by the divergent objective functions of critical stakeholder groups. Understanding these perspectives is essential for evaluating market development, infrastructure requirements, and long-term feasibility.

\subsubsection{Regulatory and Public-Sector Actors}
Regulators and central banks evaluate RWA systems primarily through the lenses of prudential oversight and systemic stability. 
Their central concern is the regulatory perimeter expecting that tokenized assets do not facilitate regulatory arbitrage by bypassing capital requirements or Know-Your-Customer (KYC) mandates \cite{fsb2023defi}. 
Consequently, bodies such as the Basel Committee, i.e., the international standard-setting authority for bank regulation and supervisory practices, have established strict risk-weighting standards (e.g., Group 1: category covering tokenized traditional assets with legal enforceability and stabilization mechanisms, versus Group 2: assets, which include unbacked or permissionless cryptoassets subject to significantly higher capital charges), to penalize banks holding unbacked or permissionless digital assets while incentivizing the use of regulated, stabilized RWAs \cite{bcbs2022prudential}. 
Furthermore, in jurisdictions of European Union, the implementation of the Markets in Crypto-Assets (MiCA) regulation seeks to create a standardized taxonomy for Asset-Reference Tokens (ARTs), mandating strict reserve segregation and redemption rights to protect investor integrity \cite{eu2023mica}.

\subsubsection{Traditional Financial Institutions}
In terms of global banks, asset managers, and custodians, the adoption of tokenization is driven by operational efficiency and collateral mobility rather than ideological decentralization \cite{jpm2022institutional}. 
Institutions view RWAs as a mechanism to reduce the balance sheet drag caused by T+2 settlement cycles, i.e., the standard two-business-day delay between trade execution and final settlement, and trapped liquidity in correspondent banking networks \cite{jpm2022institutional}.
The focus is on building permissioned or semi-permissioned environments (e.g., J.P. Morgan's Onyx, rebranded to Kinexys \cite{morgancoin}) where tokenized treasuries or money market fund shares can be used as intraday collateral for repo transactions, thereby optimizing liquidity management and automating corporate actions via smart contracts \cite{auer2023crypto}.

\subsubsection{Retail and Non-Institutional Participants}
Retail demand for RWAs is characterized by a search for yield access and portfolio diversification. 
Historically, high-grade instruments like U.S. Treasury bills, investment-grade corporate bonds, or private credit tranches were inaccessible to retail investors due to high minimum investment thresholds and geographic gating \cite{bain2023pe, wef2023private}.
Tokenization protocols enable the democratization of yield by fractionalizing these assets into accessible units, allowing retail participants to hedge against inflation using institutional-grade products previously reserved for accredited investors \cite{mckinsey2023wealth}. 
However, this demographic faces the highest risk regarding smart contract exploits and lack of recourse, necessitating proper consumer protection frameworks \cite{iosco2023defi}.

\subsubsection{Region Contexts}
In emerging economies characterized by high inflation, currency devaluation, or restrictive capital controls, tokenized assets serve a fundamentally different utility: capital preservation. 
Here, RWA adoption is driven by the demand for hard currency exposure (e.g., tokenized USD stablecoin, tokenized USD Treasuries) as a hedge against local monetary instability \cite{chainalysis2023geography}. 
In these regions, tokenized assets often function less as speculative investments and more as alternative savings vehicles or parallel payment rails, circumventing inefficient domestic banking infrastructure and lowering the cost of cross-border remittances \cite{imf2023defi}.

\section{Infrastructure}
\begin{figure*}[t]
\centering
\resizebox{\textwidth}{!}{%
\begin{tikzpicture}[
font=\sffamily\scriptsize,
assetNode/.style={draw=AssetGreen,fill=AssetGreen!10,thick,rounded corners=5pt,align=center,minimum width=2.6cm,minimum height=1.0cm,blur shadow},
legalNode/.style={draw=LegalNavy,fill=LegalNavy!8,thick,rounded corners=5pt,align=center,minimum width=2.6cm,minimum height=1.0cm,blur shadow},
bridgeNode/.style={draw=BridgePurple,fill=BridgePurple!10,thick,rounded corners=5pt,align=center,minimum width=2.6cm,minimum height=1.0cm,blur shadow},
techNode/.style={draw=TechTeal,fill=TechTeal!10,thick,rounded corners=5pt,align=center,minimum width=2.6cm,minimum height=1.0cm,blur shadow},
zone/.style={draw=black!10,fill=ZoneGray,rounded corners=8pt,inner sep=10pt},
zoneLabel/.style={font=\sffamily\bfseries\tiny,text=black!60,anchor=north west,inner sep=2pt},
flowLine/.style={draw=black!70,thick,->,>={Stealth[length=2.5mm]},rounded corners=5pt},
dataLine/.style={draw=BridgePurple,thick,densely dotted,->,>={Stealth[length=2.5mm]},rounded corners=5pt}
]
\node[assetNode](originator){Asset Originator\\{\scriptsize(Lender / Real Estate)}};
\node[legalNode,right=1.8cm of originator](spv){Special Purpose\\Vehicle (SPV)};
\node[legalNode,below=1.0cm of spv](custodian){Qualified\\Custodian};
\node[legalNode,below=1.0cm of custodian,fill=white!90!gray,text=gray!50!black,draw=gray!40](auditor){Auditor\\{\scriptsize(Proof of Reserves)}};
\node[bridgeNode,right=2.0cm of spv](engine){Tokenization\\Platform};
\node[bridgeNode,below=1.0cm of engine](oracle){Oracle Network\\{\scriptsize(Chainlink/Pyth)}};
\node[techNode,right=2.0cm of engine](token){RWA Token\\{\scriptsize(ERC-1400 / ERC-20)}};
\node[techNode,above=1.0cm of token](identity){Identity Registry\\{\scriptsize(KYC/AML Whitelist)}};
\node[techNode,below=1.0cm of token](defi){DeFi Protocol\\{\scriptsize(Lending / AMM)}};
\node[techNode,right=1.5cm of token](investor){Investor\\Wallet};
\node[legalNode,below=1.0cm of originator](docs){Legal Docs\\{\scriptsize(Trust/Indenture)}};
\node[bridgeNode,below=1.0cm of oracle](compliance){Compliance Gate\\{\scriptsize(Reg D / MiCA)}};
\node[techNode,below=1.0cm of defi](vault){Vault\\{\scriptsize(ERC-4626 / ERC-7540)}};
\node[techNode,below=0.6cm of vault](governance){Governance\\{\scriptsize(Risk Params)}};
\begin{scope}[on background layer]
\node[zone,fit=(originator)(spv)(custodian)(auditor)(docs)](legalZone){};
\node[zoneLabel]at(legalZone.north west){OFF-CHAIN LEGAL JURISDICTION};
\node[zone,fit=(engine)(oracle)(compliance)](bridgeZone){};
\node[zoneLabel]at(bridgeZone.north west){MIDDLEWARE BRIDGE};
\node[zone,fit=(token)(identity)(defi)(investor)(vault)(governance)](chainZone){};
\node[zoneLabel]at(chainZone.north west){ON-CHAIN SETTLEMENT};
\end{scope}
\draw[flowLine](originator)--node[above]{Asset Transfer}(spv);
\draw[flowLine,dashed](spv)--node[right]{Custody Agreement}(custodian);
\draw[flowLine,dashed](custodian)--node[right]{Verification}(auditor);
\draw[dataLine](auditor.east)--++(0.5,0)|-node[above,pos=0.75]{Audit Report}(oracle.west);
\draw[flowLine](spv)--node[above]{Mint Request}(engine);
\draw[dataLine](oracle)--node[right]{Data Feeds}(engine);
\draw[flowLine](engine)--node[above]{Mint}(token);
\draw[flowLine](token.north)--node[left]{Verify}(identity.south);
\draw[flowLine](token)--node[left]{Collateral}(defi);
\draw[flowLine](investor)--node[above]{Subscribe}(token);
\draw[dataLine,->] (docs.east)--++(0.8,0) |- node[right,pos=0.22,yshift=2pt]{Terms} ($(engine.north)+(0,0.1)$) -- (engine.north);
\draw[dataLine](oracle)--node[right]{Attestations}(compliance);
\draw[dataLine](compliance.east)--++(0.9,0)|-node[right,pos=0.25,xshift=2pt]{Policy}(token.south);
\draw[flowLine](defi.south)--node[left,xshift=2pt]{Deposit/Withdraw}(vault.north);
\draw[dataLine](vault.north)--++(0.2,0)|-node[right,pos=0.25,xshift=2pt]{Requests/Claims}(defi.south);
\draw[flowLine] (defi.east)--++(0.2,0) -| node[left,pos=0.55]{Yield} (investor.south);
\draw[dataLine](governance.east)--++(1.0,0)|-node[right,pos=0.25,xshift=2pt]{Update Params}(token.south);
\end{tikzpicture}}
\caption{RWA Infrastructure architecture. Solid arrows represent legal, financial, and transactional flows, while dotted arrows denote off-chain data, attestations, and control signals used for synchronization and policy enforcement. Assets are transferred from the originator to a SPV, held by a qualified custodian, and verified by auditors via proof-of-reserves. A middleware layer consisting of a tokenization platform, oracle network, and compliance gate synchronizes off-chain asset state, valuation, and regulatory constraints before minting compliant RWA tokens on-chain. On-chain settlement integrates identity verification, DeFi protocols, vaults, and governance for risk parameter updates.}
\label{fig:rwa-architecture}
\end{figure*}
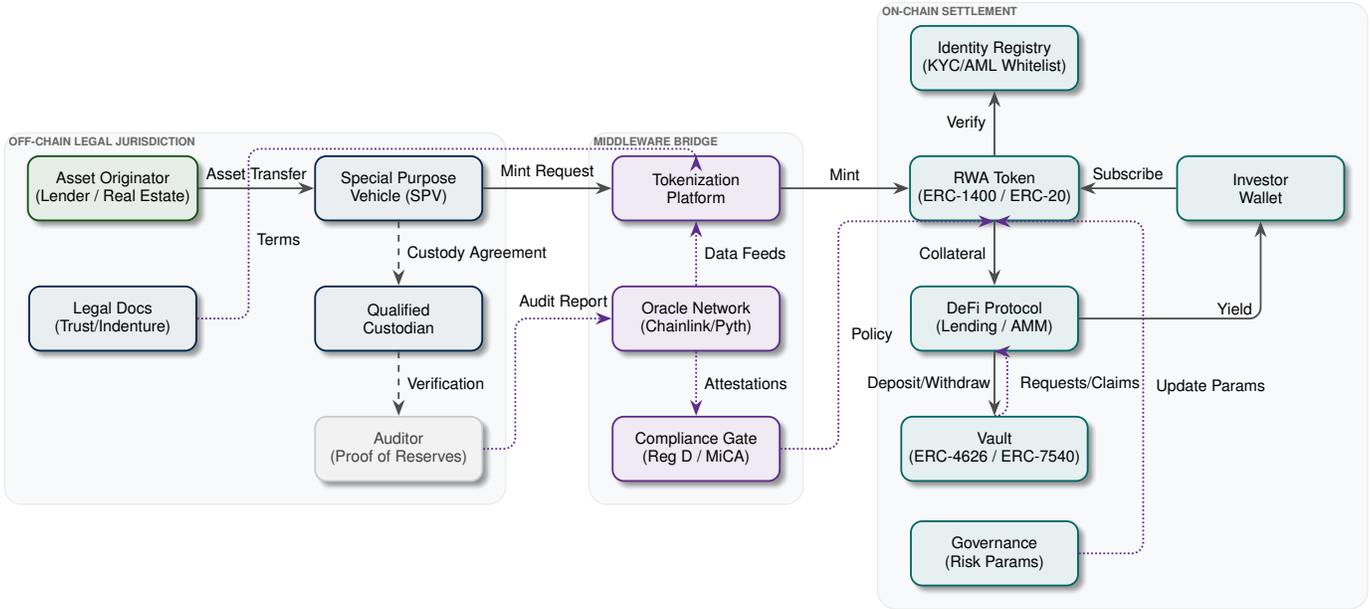

The operational efficacy of RWA systems relies on an infrastructure that integrates legal enforceability with cryptographic verification. 
The infrastructure is a composite of layers such as legal, technical, compliance, and financial, each serving a specific validation role in preserving the asset's integrity.
Figure~\ref{fig:rwa-architecture} presents an overview of the RWA infrastructure architecture, with components detailed in the following subsections.

\subsection{Legal Layer: Asset Ownership and Custody}
The foundational layer of any RWA architecture is the jurisdictional anchor that binds the on-chain token to the off-chain asset. 
Since blockchains cannot inherently enforce property rights in the physical world, this layer relies on bankruptcy remoteness, i.e., legal isolation ensuring assets are not seizable by the originator's creditors and perfection of security interests, i.e., the establishment of a priority claim over collateral against third parties to ensure token holders have a valid claim in the event of issuer insolvency \cite{schwarcz1994alchemy, cohney2019coin}.

\subsubsection{Special Purpose Vehicle (SPV)}
To isolate risk, assets are typically housed within a Special Purpose Vehicle (SPV) or a trust structure. 
The SPV functions as the legal owner of the physical asset (e.g., the deed to a building or the treasury bill account) and issues the tokens as a representation of equity or debt obligations. 
This structure is critical for bankruptcy remoteness: ensuring that if the parent company (the tokenizer) fails, the assets in the SPV remain protected and claimable by the token holders, rather than being liquidated to pay the parent company's general creditors \cite{gorton2007spv}.

\subsubsection{Qualified Custody and Asset Perfection}
Under regulatory frameworks like the U.S. SEC's Custody Rule (Rule 206(4)-2) \cite{sec2009custody}, the physical assets must be held by a qualified custodian (e.g., a regulated bank or trust company) distinct from the protocol developers \cite{sec2023custody}. 
The legal challenge lies in perfecting the token holder's interest, maintaining that the digital transfer of a token legally constitutes a transfer of the underlying asset rights without requiring a paper notarization for every secondary market trade. 
Recent innovations in commercial law, such as the adoption of Controllable Electronic Records (CERs) in the Uniform Commercial Code (UCC), i.e., the primary body of commercial law governing secured transactions and negotiable instruments in the United States, target to grant tokens the same legal standing as physical negotiable instruments, i.e., legally transferable instruments such as checks, promissory notes, or bills of exchange \cite{sec2023custody}.

\subsection{Technical Layer: Tokenization and Oracles}
The technical layer serves as the computational base, translating the legal rights defined above into smart contract logic. 
This involves two main components: the token standard governing transferability and the oracle network maintaining the data  (e.g., asset price feeds, ownership and custody status, compliance attestations) records synchronization.
While the permissive ERC-20 standard serves as the general base standard for tokens, it lacks the capability to enforce the regulatory logic required for RWA tokens (e.g., investor whitelisting, transfer restrictions, and seizure). 
Consequently, the RWA sector has evolved a diverse suite of specialized token standards designed to embed compliance, yield distribution, and asynchronous settlement.
Table \ref{tab:rwa_standards} provides a comparative overview of the standards, as detailed below:

\begin{table*}[h]
\centering
\caption{Comparative Taxonomy of RWA Token Standards}
\label{tab:rwa_standards}
\renewcommand{\arraystretch}{1.4}
\begin{tabularx}{\textwidth}{@{}l l X l@{}}
\toprule
\textbf{Standard} & \textbf{Category} & \textbf{Primary RWA Utility \& Mechanism} & \textbf{Ref.} \\
\midrule
\textbf{ERC-20} & Fungible & \textit{Base Token Functions.} The universal wrapper target, requires external whitelist contracts to restrict transfers; lacks native compliance logic. & \cite{erc20eip} \\
\midrule
\textbf{ERC-721} & Non-Fungible & \textit{Unique Asset Tokens.} The universal non-fungible tokens (NFTs). Each token has distinct identity and ownership tracking (e.g., deeds, collectibles, tokenized property), enabling metadata and unique rights per asset. & \cite{erc721eip} \\
\midrule
\textbf{ERC-1400} & Security & \textit{Partitions \& Documents.} Introduces ``partitions'' (tranches) within a single token contract and attaches legal documentation hashes to transfers. & \cite{erc1400whitepaper} \\
\midrule
\textbf{ERC-3643} & Compliance & \textit{Identity Registry.} Decouples validation logic from the token. Queries an on-chain Identity Registry (ONCHAINID) before every transfer to ensure KYC/AML compliance. & \cite{erc3643whitepaper} \\
\midrule
\textbf{ERC-4626} & Yield/Vault & \textit{Tokenized Vaults.} Standardizes the interface for yield-bearing assets (e.g., Treasury Bills). Allows RWA tokens to be instantly composable with DeFi lending protocols. & \cite{erc4626standard} \\
\midrule
\textbf{ERC-7540} & Settlement & \textit{Asynchronous Vaults.} Extends ERC-4626 to handle Request/Redeem flows. Essential for assets with settlement delays. & \cite{erc7540standard} \\
\midrule
\textbf{ERC-3525} & Semi-Fungible & \textit{Slot-Based Value.} Ideal for bonds or coupons. Allows tokens to have a unique ID (like an NFT) but hold a fungible ``value'' balance that can be split/merged. & \cite{erc3525standard} \\
\midrule
\textbf{ERC-7092} & Instrument & \textit{Financial Bonds.} Specifically designed for fixed income. Defines parameters like maturity date, coupon rate, and principal directly in the standard interface. & \cite{erc7092standard} \\
\bottomrule
\end{tabularx}
\end{table*}

\subsubsection{Compliance and Identity (ERC-3643 vs. ERC-1400)}
To address the regulatory inadequacies of the fungible ERC-20 standard, ERC-1400 introduces a security token standard based on partitioned balances, enabling a single contract to manage distinct asset tranches (e.g., restricted vs. free-trading shares) with granular control \cite{erc1400whitepaper}. 
This architecture enforces compliance by mandating that every transfer be accompanied by specific metadata (e.g., legal documentation hashes), thereby creating an on-chain audit trail absent in primitive utility tokens. 
However, the dominant compliance frameworks have shifted toward a modular permissioned model.
ERC-3643 (T-REX) has emerged as the industry standard due to its modular architecture, which separates the token contract from the identity registry. 
Unlike ERC-1400, which embeds partition logic, ERC-3643 simply validates an \textit{isVerified} boolean flag against a trusted identity oracle before executing transfers \cite{erc3643whitepaper}. 
This facilitates that if an investor's ID expires off-chain, their on-chain ability to receive tokens is automatically paused without requiring a token contract upgrade.

\subsubsection{Yield and Asynchrony (ERC-4626 vs. ERC-7540)}
For assets generating yield (e.g., tokenized U.S. Treasuries), protocols increasingly adopt ERC-4626 to ensure composability with DeFi aggregators \cite{erc4626standard}. 
Formally designated as the \textit{Tokenized Vault Standard}, ERC-4626 abstracts the yield-bearing strategies into a unified API, allowing protocols to programmatically deposit base assets and mint share tokens that represent a pro-rata claim on the underlying liquidity pool.
However, a critical friction point is the settlement mismatch between instant blockchain transactions and T+1/T+2 traditional banking wires (i.e., trade settlement occurring one ,or two business days after the trade date, respectively).
To resolve this, the ERC-7540 standard introduces asynchronous vaults, which replace atomic minting with a \textit{Request for Deposit} and \textit{Claim} workflow. 
This allows the protocol to queue investor capital on-chain, wait for the fiat wire to clear at the custodian bank, and then asynchronously mint the corresponding tokens, eliminating the risk of unbacked issuance.

\subsubsection{Oracle Networks and State Synchronization}
The fundamental constraint of blockchains is their inability to natively ingest off-chain data, creating a problem where the on-chain representation of an asset risks decoupling from its off-chain physical state. 
In RWA architectures, Decentralized Oracle Networks (DONs) function not merely as data conduits but as consensus-based truth layers \cite{al2020trustworthy}, preserving state isomorphism, the property where the digital token's parameters (e.g., supply, valuation) remain mathematically synchronized with the underlying physical collateral.
To achieve this synchronization across diverse asset classes, protocols employ a mix of oracle topologies, ranging from push-based feeds for high-frequency equivalence to optimistic assertions for subjective dispute resolution (see Table \ref{tab:rwa_oracles} in Appendix).

\paragraph{Cryptographic Solvency via Proof of Reserve (PoR)}
For fiat-backed and commodity-backed assets, the primary risk of the tokenization model is the issuance–collateral mismatch, i.e., a breach of solvency through fractional reserve issuance, which necessitates external oracle verification of off-chain reserves.
Solvency means the ability of the token issuer to fully back all issued tokens with verifiable off-chain assets, such that every token can be redeemed one-for-one without shortfall.
Chainlink Proof of Reserve (PoR) mitigates the risk by establishing an autonomous circuit breaker: oracle nodes periodically query the custodian's off-chain API to verify the exact balance of assets held in the SPV (a legally separate entity created to hold the underlying assets and isolate them from the issuer’s balance sheet) \cite{breidenbach2021chainlink}. 
This data is cryptographically attested on-chain, allowing the token contract to enforce atomic solvency constraints, automatically reverting any mint transaction that would cause the total circulating supply to exceed the verified off-chain reserves \cite{breidenbach2021chainlink}.

\paragraph{Alternative Topologies: Latency, Liability, and Cost}
Beyond solvency, distinct RWA sub-sectors necessitate specialized oracle architectures. 
For market-correlated RWAs (e.g., synthetic equities), protocols prioritize the high-frequency update cycles of the Pyth Network (A decentralized oracle network that delivers real-time market prices directly from trading venues to blockchains), which aggregates first-party publisher data to minimize the latency arbitrage risks in slower push-based models \cite{pyth2024whitepaper}. 
Conversely, institutional issuers facing strict liability constraints (e.g., the European Union’s General Data Protection Regulation (GDPR) governing data protection and privacy) favor API3's first-party Airnode architecture, which eliminates third-party node operators to create a legally accountable cryptographic tunnel between the banking ledger and the smart contract \cite{api32020whitepaper}. 
Finally, to address the storage costs of long-tail assets (e.g., diverse carbon credit vintages), RedStone implements a modular lazy-loading design, injecting signed data into the EVM memory stack only during transaction execution to bypass expensive on-chain storage operations \cite{redstone2023docs}.

\paragraph{Subjective State Verification (Optimistic Oracles)}
While objective data can be cryptographically proven, subjective RWA state, e.g., "Has the real estate property suffered flood damage?" require a human-in-the-loop consensus mechanism. 
Protocols like UMA \cite{uma2018whitepaper} utilize an optimistic assertion model, where a state update is proposed on-chain and assumed valid unless disputed within a liveness window by a bonded actor. 
This mechanism introduces a necessary trade-off of immediate finality in exchange for the temporal capacity to accurately adjudicate off-chain states.

\subsection{Compliance Layer: Identity and Regulation}
Real World Assets operate within strict regulatory perimeters (e.g., the U.S. Securities and Exchange Commission's Regulation D, specifically Rule 506(c), which mandates rigorous verification of accredited investor status for private capital formations \cite{sec2013regd}, and the European Union's Markets in Crypto-Assets (MiCA) Regulation, which imposes comprehensive disclosure, capital reserve, and licensing requirements on issuers of asset-referenced tokens \cite{eu2023mica}).
the compliance layer functions as a gatekeeping middleware, preserving compliance by design, where regulation is not mere a legal agreement but a deterministic constraint on transaction validity.

\subsubsection{Decentralized Identity (DID) and Verifiable Credentials}
The primitive method of compliance, maintaining a static whitelist of addresses in a smart contract, suffers from scalability and privacy defects (i.e., revealing the entire investor list on-chain). 
Modern RWA architectures instead adopt the Verifiable Credential (VC) model standardized by the W3C \cite{w3c2019vc, w3c2025vc}.
In this architecture, an investor undergoes KYC off-chain with a regulated provider. 
The provider issues a cryptographically signed VC (attesting to attributes such as accredited investor, i.e., an individual or entity meeting regulatory income, net worth, or sophistication thresholds, or non-sanctioned, i.e., not listed on government sanctions or prohibited-party registries) to the user's wallet. 
The token smart contract does not store user data; instead, it utilizes an on-chain identity registry to cryptographically verify the validity of the VC signature before permitting any token transfer. 
This decouples identity verification off-chain from access authorization on-chain, maintaining GDPR compliance by keeping Personally Identifiable Information (PII) off the public blockchain.

\subsubsection{Privacy-Preserving Compliance (zkKYC)}
To resolve the tension between public auditability and investor privacy, protocols increasingly integrate zero-knowledge proofs (ZKPs). 
Utilizing schemes like zk-SNARKs, an investor can generate a proof that satisfies a predicate (e.g., age and country not in a sanctioned list) without revealing the underlying values or their real-world identity to the verifier \cite{garman2016credentials}. 
This allows RWA platforms to prove compliance to regulators (via the proof) while maintaining the pseudonymity of the blockchains.

\subsubsection{Regulatory Enforcement: The Travel Rule}
Beyond initial issuance, the compliance layer must preserve the regulatory mandates such as the FATF Travel Rule, which requires Virtual Asset Service Providers (VASPs) to exchange originator and beneficiary information for transactions exceeding a certain threshold \cite{fatf2021guidance}. 
RWA contracts implement this via transaction hooks: before a transfer settles, the contract checks if the receiving address belongs to a compliant VASP or a verified self-hosted wallet. 
If the counterparty risk assessment fails (e.g., the address is tainted by mixer interaction), the smart contract atomically reverts the transaction, thereby embedding Anti-Money Laundering (AML) controls directly into the asset's execution logic.

\subsection{Financial Layer: Liquidity and Credit}
While the legal and technical layers establish the asset's existence, the financial layer activates the utility. 
The layer integrates tokenized assets into DeFi market structures, fundamentally transforming illiquid instruments (e.g., real estate equity) into productive collateral. 
We demonstrate the utility by four primary primitives: Automated Market Makers (AMMs) for exchange, over-collateralized lending for leverage, structured tranching for risk redistribution, and asynchronous liquidation modules for default management.

\subsubsection{Concentrated Liquidity and Swap}
Traditional order book models work for RWAs but could be insufficient, which generally exhibit low velocity and thin trading volumes. RWA trading protocols also utilize concentrated liquidity AMMs (e.g., Uniswap v3). 
Concentrated liquidity allows Liquidity Providers (LPs) to allocate capital within a specific, narrow price range. 
This capital efficiency is critical for RWAs; empirical data suggests that concentrated liquidity can support trading volumes much larger than legacy constant product models (e.g., Uniswap v2) for stable-pairs, significantly reducing slippage for institutional-sized exits \cite{adams2021uniswap}.

\subsubsection{Collateralization and Isolation Markets}
The integration of RWAs into lending markets (e.g., Aave, MakerDAO) introduces unique credit risks distinct from crypto-native volatility. 
To mitigate systemic contagion, protocols employ isolation mode. 
In this mode, a risky RWA (e.g., a tokenized private credit fund) is segregated: it can be used as collateral to borrow stablecoins, but only up to a specific debt ceiling (e.g., \$10 million) and often cannot be mixed with other collateral types in the same position. 
Furthermore, the Loan-to-Value (LTV) ratios for RWAs are typically conservative compared to liquid crypto-assets, reflecting the lower liquidity and longer settlement times required to liquidate the underlying physical asset in the event of default \cite{aave2022v3}.

\subsubsection{Structured Credit and Risk Tranching}
Real-world credit portfolios, i.e., pools of loans, receivables, or debt instruments backed by borrowers or assets, are heterogeneous in risk, encompassing varying probabilities of default, loss given default, maturity profiles, jurisdictional enforcement uncertainty, and underlying borrower or asset quality across individual exposures.
To address this, protocols such as Centrifuge \cite{centrifuge2021whitepaper} utilize wateralled tranching to redistribute default risk among investors. 
The SPV typically issues two distinct token classes against the same collateral pool: a senior tranche (known as DROP, refers the waterfall payment structure, a "drop" of the safe liquidity) that prefers priority on cash flows, and a junior tranche (known as TIN, derived from the protocol name Tinlake \cite{schmitt2019centrifuge}, and possibly refers "tin" as a base metal compared to the "gold" standard of the senior tranche) that absorbs the first losses, i.e., it bears initial defaults or shortfalls in the collateral pool before any losses are allocated to the senior tranche (if some underlying loans fail to pay back, the junior tranche loses value first) \cite{centrifuge2021whitepaper}.
\begin{enumerate}
    \item \textbf{Senior Tokens (DROP):} A fixed-yield, low-risk instrument protected by a "first-loss" buffer.
    \item \textbf{Junior Tokens (TIN):} A variable-yield, high-risk instrument that absorbs the first defaults in the portfolio.
\end{enumerate}
This structure mimics traditional securitization (CLOs), i.e., financial structures that pool loans and redistribute risk across senior and junior tranches with different return and risk profiles.
Smart contracts enforce an algorithmic waterfall: borrowers repay the pool, and the contract fills the senior yield buckets first.
Only after seniors are fully paid does the remaining capital flow to juniors. 
This mechanism allows RWA protocols to crowd-source junior capital from risk-seeking DeFi traders to insure the Senior capital provided by conservative institutional treasuries \cite{centrifuge2021whitepaper}.

\subsubsection{Liquidation Latency and Insolvency Buffers}
The most significant friction in the financial layer is the liquidation mismatch. 
In crypto-native DeFi, a liquidation is atomic: if ETH price drops, the protocol sells it on Uniswap in the same block. 
For RWAs, physical liquidation is a legal process that can take months (e.g., foreclosing on a property).

To manage this, protocols such as MakerDAO \cite{makerdaowhitepaper} utilize a \textit{Tell-and-Cure} oracle mechanism rather than an atomic auction.
\begin{enumerate}
    \item \textbf{Tell (Trigger):} The trustee reports a default event on-chain.
    \item \textbf{Cure (Grace Period):} The protocol freezes the value of the collateral and grants a grace period (e.g., 24 hours) for the borrower to resolve the issue.
    \item \textbf{Write-Off:} If the cure fails, the protocol does not attempt to sell the token (which is now illiquid). Instead, it triggers a \textit{Debt Write-Off}, absorbing the loss via the protocol's Surplus Buffer (insurance fund) while the legal team pursues the physical assets off-chain.
\end{enumerate}
This architecture, the decoupling of immediate on-chain debt cancellation from delayed off-chain enforcement, prevents a fire sale spiral, i.e., a self-reinforcing price crash where forced liquidations overwhelm market liquidity. 
By absorbing the immediate loss via the surplus buffer, the protocol circumvents the need to dump the underlying assets into an illiquid market (similar to a bank using its own capital to cover a bad loan while waiting for a foreclosure auction to maximize the sale price), thereby preserving the collateral's realization value during the lengthy legal recovery process \cite{maker2021mip21}.

\subsection{Governance Layer: Protocol and Risk Management}
While compliance focuses on external legality, e.g., "Is this user allowed to trade?", governance focuses on internal risk parameterization, e.g., "Should the protocol accept this asset?". 
This layer is the decision-making engine that updates the smart contract parameters based on evolving economic conditions.

\subsubsection{Decentralized Risk Assessment (DAO Governance)}
In RWA protocols, the Decentralized Autonomous Organization (DAO) replaces the traditional credit committee. 
Token holders vote on critical risk parameters, including the stability fee (interest rate charged to borrowers), the liquidation threshold (the price point at which collateral is seized), and the oracle security module delay. 
This process involves a real world assessment. 
For instance, before MakerDAO onboards a new RWA vault, it requires a legal assessment of the SPV structure to ensure bankruptcy remoteness and a credit assessment of the underlying borrower's cash flows \cite{maker2021rwa}. 
Unlike code updates, these decisions are subjective and rely on delegate models, where token holders delegate voting power to professional risk units who interpret the off-chain legal data and translate it into on-chain bytecode execution.

\subsubsection{Emergency Shutdown and Pause Modules}
A unique feature of RWA governance is the Emergency Shutdown Module (ESM). 
Given that physical assets are subject to seizure, freeze orders, or destruction (e.g., natural disasters affecting real estate), governance contracts often include a pause functionality. 
The functionality allows a security council (elected multisig) to instantly freeze protocol interactions if the oracle reports a de-peg event or if the off-chain custodian reports a regulatory breach. 
This kill switch mechanism is a necessary centralization trade-off to protect solvency in a system bridging trustless code with trust-based physical laws.

\subsection{Cross-Layer Coordination Mechanisms}
The primary architectural challenge in RWA systems is the dissonance between deterministic code and probabilistic real-world states. 
The functional integrity of the system relies on coordination loops that force the distinct layers: legal, technical, financial, compliance, and governance, to operate as a synchronous unit. 
We define two primary primitives that bridge these components: 

\subsubsection{Algorithmic Solvency Synchronization}
This mechanism coordinates the legal layer (custody) with the financial layer (token supply) using the technical layer (oracles) as the binding agent. 
A fundamental risk in RWAs is the decoupling of liability from asset, where the on-chain financial supply potentially exceeds the off-chain legal collateral.
To prevent this, protocols implement a concept known as embedded supervision \cite{auer2022embedded}. 
The architecture creates a closed-loop dependency:
\begin{enumerate}
    \item \textbf{Legal State:} The qualified custodian exposes a read-only API reporting the real-time verified balance of the SPV's vault.
    \item \textbf{Technical Bridge:} A Proof of Reserve (PoR) oracle node continually queries this API and cryptographically attests the value on-chain.
    \item \textbf{Financial Execution:} The token's smart contract enforces a hard logic gate: the minting cap equals the oracle reserve.
\end{enumerate}
The architecture preserves that the financial layer is technically incapable of issuing unbacked liabilities, so that the digital token remains mathematically isomorphic to the physical asset held in the legal layer without manual human intervention.

\subsubsection{Governance-Mediated Risk Translation}
While code handles data such as prices, code cannot fully interpret subjective legal risks, e.g., "Is this jurisdiction enforcing bankruptcy laws?".
This coordination mechanism uses the governance layer to translate legal features into financial parameters.
As analyzed by prior studies \cite{zetzsche2020defi}, this requires a parameterization pipeline that bridges the off-chain and on-chain worlds:
\begin{enumerate}
    \item \textbf{Legal Input:} Independent auditors assess the SPV's legal structure, analyzing the enforceability of liens and the credit quality of the underlying borrowers.
    \item \textbf{Governance Translation:} Token holders vote to quantify this qualitative legal risk into numerical scalars of the collateralization ratio and the liquidation threshold.
    \item \textbf{Financial Output:} These parameters are hardcoded into the lending vaults.
\end{enumerate}

\section{Technical Challenges and System Constraints}
Despite the modular efficacy of the infrastructure layers described, the integration of physical assets into blockchains introduces severe gaps of technical challenges \cite{caldarelli2020oracle}.
These gaps manifest not merely as an implementation hurdle but as a fundamental limitation in systems, where the security of the asset is no longer defined solely by cryptographic hardness but by the weakest link in the off-chain dependency chain \cite{bis2023oracle}.
Consequently, current RWA architectures face critical bottlenecks in data freshness, contract immutability, and privacy preservation that limit their scalability from experimental pilots to systemic financial infrastructure \cite{eskandari2021sok}.
The limits and challenges are summarized as follows.

\subsection{Oracle and Data Latency}
One of the most acute vulnerability in RWA systems is the reliance on asynchronous data injection, which breaks the atomicity of the blockchain state. 
Unlike native tokens that update instantly within a block, real-world data (e.g., the appraisal value of a property or the audit status of a gold vault) suffers from non-zero ingestion latency (i.e., the time lag between the physical event and its on-chain registration.
This latency creates a temporal arbitrage window exploitable by adversarial actors via Maximal Extractable Value (MEV) strategies \cite{daian2020flash}. 
If an off-chain asset price changes significantly but the oracle update is pending in the mempool (i.e., the waiting area for unconfirmed transactions), automatic bots can front-run the update to trade against the stale on-chain price, effectively stealing value from the protocol's liquidity providers before the smart contract knows the true price \cite{daian2020flash}.
Furthermore, the integrity of this data pipeline is often compromised by source-level centralization, i.e., reliance on a single off-chain authority or data provider as the origin of truth. 
While the delivery network (e.g., Chainlink) may be decentralized, the original data source (e.g., a single appraiser's API or a specific bank's ledger) acts as a Single Point of Failure (SPOF). 
If this solitary source is corrupted or goes offline, the decentralized network merely achieves consensus on false data (erroneous inputs deterministically propagate into erroneous on-chain outcomes), leading to incorrect liquidations or unauthorized minting that the blockchain cannot cryptographically detect \cite{eskandari2021sok}.

\subsection{Smart Contract Security}
Tokenization, custody, and distribution logic rely on contract architectures that remain vulnerable to implementation bugs and upgrade risks \cite{zhou2022sok}.
Unlike permissionless DeFi primitives, which are immutable upon deployment, RWA protocols must retain mutability to address changing regulatory landscapes (e.g., updating a blacklist to comply with new U.S. Office of Foreign Assets Control (OFAC) sanctions) \cite{min2022upgradable}.
This necessity introduces the upgradeability. 
To allow legal updates, protocols employ proxy patterns, i.e., architecture where a stable proxy contract delegates execution to a mutable logic contract. 
However, this introduces severe security fragility: storage layout collisions i.e., when a new logic contract accidentally overwrites the variables of the old one, can permanently corrupt the asset's ledger, rendering the tokenized equity unclaimable \cite{min2022upgradable, trailofbits2020upgrade}. 
Also, the existence of an upgrade mechanism may imply the existence of an admin key (i.e., a privileged cryptographic key capable of altering the contract's logic). 
If this key is compromised via phishing or social engineering, an attacker can perform a logic rug pull, arbitrarily upgrading the contract to a malicious version for unlimited minting or confiscation of user funds \cite{zhou2022sok}.
Beyond governance risks, the implementation of compliance hooks (e.g., ERC-3643 transfer checks) expands the attack surface for reentrancy attacks i.e., attacks in which a malicious contract repeatedly re-enters a function before its previous execution completes, exploiting intermediate states to manipulate balances or control flow.
Since RWA tokens often trigger external calls to identity registries during every transfer, they violate the checks-effects interactions pattern (i.e., the security  practice of updating internal state variables before executing external calls to prevent reentrancy) \cite{rodler2018sereum}.
A malicious registry or a compromised adapter could potentially hijack the control flow during the callback, recursively calling the transfer function to drain the lending pool before the balance is updated \cite{rodler2018sereum}, therefore undermining the solvency and safety guarantees of the protocol.
The complexity of Role-Based Access Control (RBAC) creates vulnerabilities in privilege management. 
RWA contracts often contain dozens of distinct roles (e.g., \textit{Minter}, \textit{Burner}, \textit{Pauser}, \textit{ComplianceOracle}) \cite{ferreira2020access}. 
Empirical analysis suggests that as the number of privileged roles increases, the probability of privilege escalation bugs, i.e., a regular user tricking the contract into granting them admin rights grows non-linearly, commonly due to the errors in modifier logic utilized across modular components \cite{perez2021vulnerabilities, torres2021frontrunner}.

\subsection{Scalability and Cost Constraints}
The economic viability of RWA protocols is strictly bounded by the computational overhead and throughput saturation of the underlying ledger.
High-fidelity assets (e.g., rental real estate or trade finance invoices) require frequent state updates to reflect cash flow distributions or credit rating changes. However, the throughput limitations of Layer-1 (L1) blockchains such as Ethereum (approx. on the order of 15 to 20 TPS) render such granular updates cost-prohibitive.
This creates a micro-transaction inefficiency: if the gas cost to distribute a monthly dividend exceeds the value of the dividend itself, the asset becomes functionally illiquid, restricting RWA adoption to high-net-worth tranches while excluding the retail investors the technology aims to empower \cite{lewenberg2015inclusive, croman2016scaling}. 
As the history of the chain grows, the system suffers from state bloat, a condition where the storage requirements for running a full node exceed consumer hardware capabilities: as of December 2025, an Ethereum full node (default/pruned sync) typically stores 1 TB of data if not aggressively pruned, yet many operators recommend provisioning 4–8 TB storage to run a full node reliably with growth headroom, while archive nodes retaining all historical state are substantially larger, often requiring 12–20 TB on Geth \cite{quicknode2025}; with client-specific differences such as Erigon \cite{erigon_github2025} 2–3 TB, and continuing to grow.
This centralization pressure forces RWA protocols to rely on third-party RPC providers (e.g., Infura \cite{infura2025}, Alchemy \cite{alchemy2025}, QuickNode \cite{quicknode2025}) to read their own data, reintroducing the intermediary risk they sought to eliminate.
To mitigate these constraints, RWA protocols increasingly adopt Layer-2 (L2) rollups (so termed because transactions are 'rolled’, like pages in a ledger bundled together, aggregated off-chain and then 'rolled up’ into a single compressed proof submitted to the base layer) as execution environments while retaining Layer-1 (L1) blockchains as the canonical settlement and registry layer \cite{thibault2022rollups}.
In this architecture, asset ownership, issuance limits, and legal anchors are recorded on L1, whereas high-frequency operations such as trading, collateral rebalancing, and cash-flow accounting are executed on L2 to reduce cost and latency.
This separation is particularly critical for RWAs, which require frequent state updates and institutional-scale throughput while still relying on L1 security guarantees for finality and legal verifiability.
However, the L2-L1 design introduces trade-offs in withdrawal latency, proof generation overhead, and cross-layer synchronization that directly affect liquidity management and settlement assurance \cite{mccorry2021sok}.

\subsection{Interoperability Boundaries and Liquidity Fragmentation}
The RWA ecosystem is currently plagued by liquidity fragmentation, where assets are isolated within incompatible execution environments. 
For instance, a tokenized bond issued on a permissioned bank ledger (e.g., Onyx \cite{jpmorgan2020onyx}) cannot be natively collateralized on a public DeFi protocol (e.g., Aave \cite{aave2020whitepaper}).
This creates closed, siloed capital structures, breaking the core DeFi promise of global composability and forcing institutions to manage fragmented liquidity pools across dozens of isolated chains \cite{pillai2022fragmentation}.
To bridge these silos, protocols utilize cross-chain bridges, but these introduce the wrapper risk paradox. 
A bridged asset is not the asset itself, but a synthetic claim on a locked asset on the source chain. This architecture creates a massive honeypot effect; historical data indicates that bridge vulnerabilities account for approximately 69\% of all funds lost in DeFi exploits, as attackers target the central storage contracts that hold the locked collateral rather than the assets themselves \cite{chainalysis2022bridge, belchior2021survey}.
Moreover, achieving atomic composability (i.e., the ability to execute a swap where both legs settle simultaneously or neither does) across asynchronous ledgers remains a theoretically unsolved distributed systems problem. 
Current solutions such as hash time-locked contracts (HTLCs) suffer from the free option problem: an adversary can lock a counterparty's funds and wait to see if the exchange rate moves in their favor before deciding to finalize or abort the trade \cite{herlihy2018atomic}. 
Consequently, most institutional RWA transfers must rely on trusted notary schemes (e.g., Multi-Sig Federations) rather than trustless code to coordinate cross-chain settlement, effectively reverting the architecture to a federated banking model \cite{zamyatin2021sok}.

\subsection{Privacy and Selective Disclosure}
A critical paradox in RWA adoption is the privacy-transparency dilemma: while public blockchains rely on radical transparency for trust, financial markets rely on confidentiality for competitive advantage. 
Institutional participants may not utilize standard tokenization such as ERC-20 because the open ledger exposes sensitive proprietary data, such as trade sizes, portfolio compositions, and counterparty relationships. 
This on-chain leakage allows adversarial actors to reverse-engineer trading strategies or front-run large institutional orders based on visible wallet movements \cite{zhang2019security, kosba2016hawk}.
The immutability of the ledger creates a direct conflict with data protection frameworks such as the General Data Protection Regulation (GDPR). 
Specifically, the right to be forgotten (Article 17 of the GDPR grants individuals the right to have their personal data erased when it is no longer necessary for the purposes for which it was collected or when consent is withdrawn \cite{gdpr2016article17}), is technically impossible to satisfy on a permanent ledger; if a user's identity or transaction history is recorded on-chain, it cannot be erased, exposing issuers to significant regulatory liability \cite{finck2019blockchains}.
%

%
To address this, recent protocols are leveraging zero-knowledge proofs and confidential payment constructions (e.g., Zether \cite{bunz2020zether}), as well as compliance-aware privacy layers for ERC-3643 (e.g., UltraMixer \cite{ultramixer}), to enable selective disclosure, i.e., proving transaction validity without revealing the sender or the transferred amount \cite{bunz2020zether,ultramixer}. 
However, embedding privacy and compliance into the execution path can impose substantial overhead: generating proofs for composite eligibility predicates (e.g., "accredited AND not sanctioned AND balance $>$ transfer”) can cost orders of magnitude more gas and prover time than a plain transfer, which in turn limits the practical throughput of the RWA systems \cite{ultramixer}.
The adoption of decentralized identifiers (DIDs) and verifiable credentials (VCs) offers a solution to the aforementioned privacy problem. 
Traditional Public Key Infrastructure (PKI) where a certificate authority observes all validation requests, while DIDs enable a self-sovereign architecture where the user retains control over their attestation data \cite{sporny2022did}. 
The DIDs' infrastructure supports selective disclosure schemes, allowing an investor to cryptographically prove the attributes (e.g., "accredited investor status" or "domicile is not US") without revealing their full identity or underlying documents to the on-chain verifier. 
Through decoupling the attestation from the transaction, protocols satisfy regulatory KYC requirements while reducing the risk of storing sensitive personal identifiable information (PII) on centralized servers \cite{meiklejohn2019coconut}.

\subsection{Exogenous Economic Risks}
\label{sec:exogenous_risks}
While technical and legal layers address operational integrity, the economic viability of RWAs is subject to broader market forces. We summary three critical exogenous risks that extend beyond the immediate scope of protocol architecture.
\begin{itemize}
    \item \textit{Macro-Financial Dependencies (Interest Rate Sensitivity):} 
    Crypto-native assets which often behave orthogonally to traditional markets, while RWAs introduce direct correlation risk. As risk-free rates (e.g., U.S. Treasury yields) rise, the opportunity cost of holding on-chain stablecoins increases, potentially triggering liquidity drains from RWA protocols as capital rotates back to traditional off-chain instruments. This yield sensitivity imports traditional monetary policy shocks into the DeFi \cite{fsb2023defi}.

    \item \textit{Systemic Concentration (Custodial Monopolies):} 
    Although protocols are decentralized, the underlying custody market exhibits high consolidation. If a single qualified custodian (e.g., BNY Mellon \cite{bnymellon2022launch} or Coinbase Custody \cite{carreras2024concentration}) secures the physical collateral for multiple dominant RWA protocols, it creates a Systemically Important Financial Institution (SIFI) failure point. Operational failures or regulatory freeze at this custodian would trigger simultaneous insolvency across unrelated RWA protocols, nullifying the diversification benefits of the asset class \cite{feyen2021fintech}.

    \item \textit{Inadequacy of Stress Testing Frameworks:} 
    Current risk models (e.g., Aave's safety module \cite{aave2020safetymodule}) account for crypto-native volatility but lack multi-asset correlation matrices for physical markets. There is an absence of standardized \textit{Value-at-Risk (VaR)} frameworks capable of simulating "dual-shock" scenarios, where a macro-economic downturn depresses real estate prices (RWA collateral) while simultaneously triggering a crypto-market sell-off (ETH collateral), leaving protocols under-collateralized during systemic crises.
\end{itemize}

\section{Legal Interoperability and Regulation}
The technical infrastructure for RWAs is approaching maturity even with constraints, but the legal layer remains immatured for systemic adoption. 
The fundamental regulatory challenge lies in ensuring legal enforceability and consistency between on-chain state and off-chain legal reality, such that ownership, transfer, collateralization, compliance status, and insolvency outcomes encoded and executed by smart contracts are recognized, enforceable, and overrideable when necessary by courts, regulators, and insolvency administrators across jurisdictions that were designed for traditional financial systems.

\subsection{Asset Ownership and Recognition Frameworks}
Unlike native cryptoassets (e.g., Bitcoin) where the token is the asset, RWAs are mostly digital representations of off-chain liabilities. 
Consequently, they rely on special purpose vehicle (SPV) or trust-based structures to bridge the digital-physical divide. 
The legal soundness of this bridge depends on the token container model, a framework pioneered in the Liechtenstein Blockchain Act (TVTG), which legally defines the token as a "container" that holds the underlying right (e.g., ownership, lien, or copyright) \cite{duenser2019liechtenstein}.
However, this structure introduces the risk of separation of title. 
If the private keys controlling the token are stolen, but the legal title to the underlying real estate remains registered to the victim in the official land registry, a legal fork occurs. 
Courts must then decide whether the blockchain record (possession) or the state registry (title) takes precedence. 
In Common Law jurisdictions, i.e., judge-made law developed through judicial precedent rather than statute \cite{dainow1966civil}), recent guidance suggests that while tokens constitute property capable of being held in trust, the enforceability of the claim against the SPV depends entirely on the robust integration of the smart contract logic into the corporate bylaws \cite{ukjt2019legal, konashevych2020token}. 
Without this integration, the token is reduced to a vanity representation with no standing in insolvency proceedings.
Recent legal developments seek to resolve the ambiguity of virtual possession by providing statutory clarity around digital asset rights. 
A pivotal advancement is the 2022 amendment to the Uniform Commercial Code (UCC) in the United States \cite{ulc2022amendments}, specifically Article 12 \cite{ucc_article12_2022}, which establishes the category of Controllable Electronic Records (CERs).
This framework legally equates control of a private key with physical "possession" allowing digital assets to benefit from the same take-free rules as negotiable instruments, meaning a purchaser in good faith takes clear title even if the seller had a defective title \cite{marcantel2024article12}.
Simultaneously, the European Union has implemented the Markets in Crypto-Assets (MiCA) regulation \cite{eu_mica_regulation_2023}, which creates a comprehensive taxonomy for Asset-Referenced Tokens (ARTs) and E-Money Tokens (EMTs). 
Unlike the U.S. approach which focuses on commercial law and perfection of security interests, MiCA focuses on prudential supervision, requiring issuers to maintain 1:1 liquid reserves and providing token holders with a statutory right of redemption against the issuer \cite{eu2023mica}. 
This regulatory unevenness affects the portability of RWA tokens; a token compliant with UCC Article 12 \cite{ucc_article12_2022} as a promissory note may legally degrade into an unregulated security when transferred to a non-MiCA jurisdiction, fragmenting the secondary market.

\subsection{Jurisdictional Fragmentation and Conflict of Laws}
The borderless nature of distributed ledgers fundamentally conflicts with the Westphalian model of territorial sovereignty, creating a phenomenon known as jurisdictional fragmentation \cite{zetzsche2018distributed, xiong2024global}. 
Different legal systems possess incompatible taxonomies for digital instruments, e.g., while Switzerland and Liechtenstein may recognize a tokenized share as a distinct uncertificated security transferable via blockchain, other jurisdictions such as Germany and Netherlands may classify the same instrument as a mere contractual promise, requiring notarized paper deeds for valid transfer \cite{hacker2018crypto}.
This divergence creates a conflict of laws crisis for secondary markets. 
If a token issued in Singapore (under Common Law) is purchased by an investor in Germany (under Civil Law), it is unclear which legal regime governs the finality of the settlement. 
This legal uncertainty fractures global liquidity pools, as institutional issuers must ring-fence assets into compliant silos, i.e., legally and operationally segregated pools of capital that cannot freely interact with other markets or jurisdictions, to avoid regulatory arbitrage or inadvertent securities violations \cite{oecd2020fragmentation}. 
To mitigate this, international bodies such as the International Institute for the Unification of Private Law (UNIDROIT) have proposed the \textit{Principles on Digital Assets and Private Law} \cite{unidroit2023principles}, advocating for an echnologically neutral private law framework where the law of the platform or the issuer's choice of law takes precedence over the physical location of the server nodes.

\subsection{Compliance, Identity, and Transfer Controls}
Legal interoperability also depends on mechanisms for identity verification and transfer restrictions, which directly conflict with the permissionless of public blockchains. 
To adhere to anti-money laundering (AML) directives, RWA protocols must enforce the Financial Action Task Force (FATF) (the intergovernmental body that sets global standards for combating money laundering and terrorist financing), Travel Rule (Recommendation 16) \cite{fatf2021guidance}, which mandates that Virtual Asset Service Providers (VASPs) exchange originator and beneficiary customer data alongside any transaction over a certain threshold \cite{fatf2021guidance}.
Since standard ERC-20 tokens cannot carry this metadata, the industry is adopting permissioned token standards such as ERC-3643 (T-REX). 
This standard introduces an on-chain identity registry that decouples the wallet address from the user's identity. 
The token contract implements a transfer with authorization function that queries a trusted identity oracle before every transaction; if the receiver's wallet does not hold a valid identity claim (e.g., KYC cleared, accredited investor status), the transaction reverts automatically. 
This architecture ensures that the asset remains strictly within the restricted participation domain of compliant participants, even if the underlying ledger is public and permissionless \cite{tokeny2023erc3643}.
%


\section{Protocols and Representative Implementations}

\label{sec:protocols}
This section surveys representative RWA implementations across the major asset classes. Table~\ref{tab:rwa-projects} in Appendix~\ref{sec:rwa-projects} provides a comparison along the lifecycle-relevant dimensions of (i) legal wrapping and custody, (ii) on-chain representation and settlement environment, and (iii) compliance, audit, and data synchronization. The goal is to show design choices concrete: how treasury-like products operationalize NAV and redemption, how private-credit protocols encode tranche and delegate risk, how real-estate systems bind localized operations to global settlement rails, and how commodity tokens reduce solvency to custody and attestation primitives. Descriptive content and comparative analysis of representative RWA protocols are deferred to Appendix~\ref{sec:rwa-projects}.

\section{Empirical Case Study}



\subsubsection{Tokenized U.S. Treasuries (GovDebt)}

We analyze the on-chain behavior of tokenized U.S. Treasury products using Ondo Finance's OUSG \cite{ondo-ousg} as a representative case study. OUSG is an ERC-20 token on Ethereum that provides exposure to short-term U.S. Treasury securities, with each token representing a fractional claim on the underlying portfolio. We collected all token transfer events from the Ethereum blockchain for the period January 2024 through December 2025, and obtained Federal Funds Effective Rate data.

\begin{figure}[t!]
    \centering
    \includegraphics[width=0.5\textwidth]{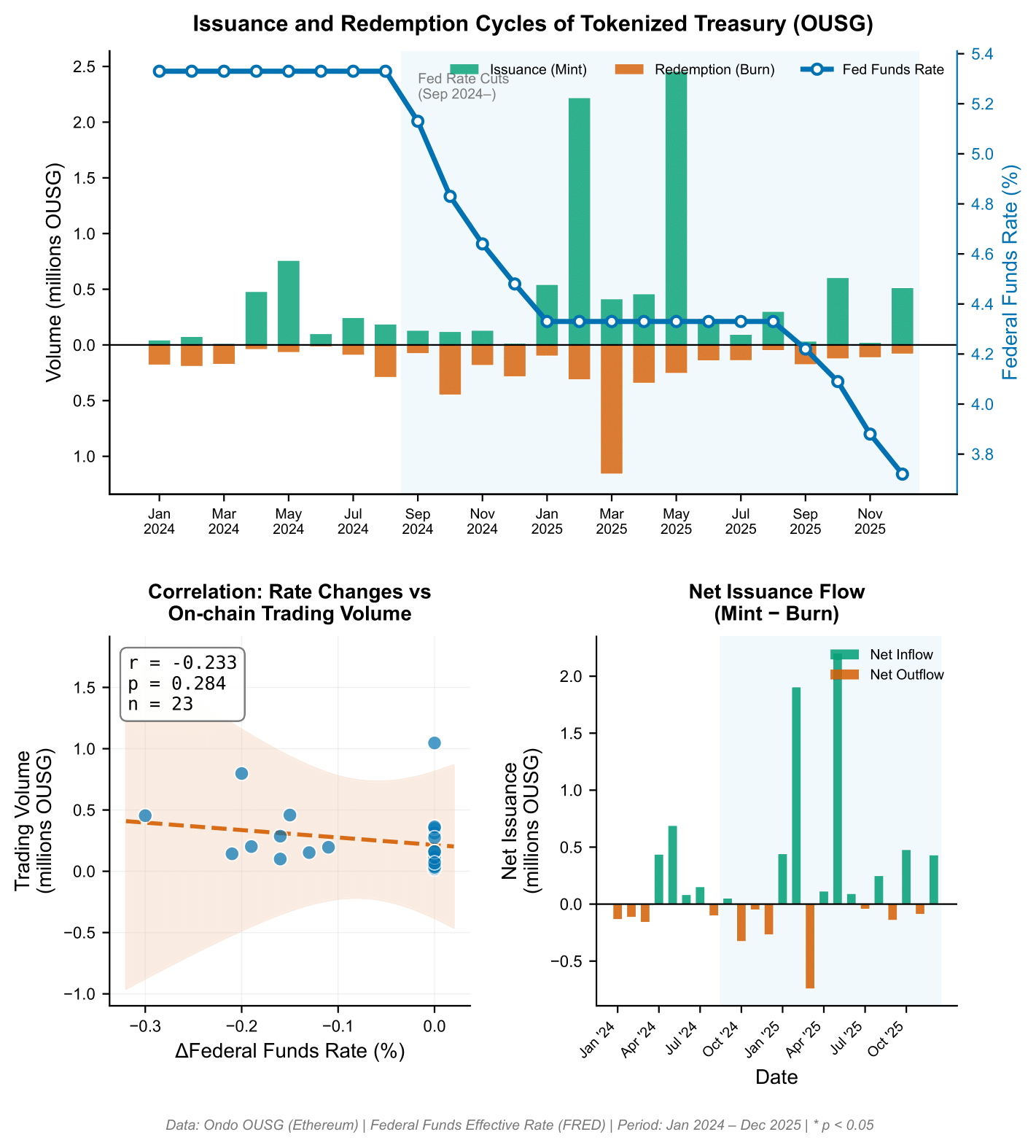}
    \caption{On-chain activity of tokenized U.S. Treasury (Ondo OUSG) and Federal Funds Rate dynamics.}
    \label{fig:govdebt}
\end{figure}

\paragraph{Issuance and Redemption Cycles}
Figure~\ref{fig:govdebt} (top panel) illustrates the monthly issuance (mint) and redemption (burn) volumes of OUSG tokens alongside the Federal Funds Rate trajectory. During the study period, the Federal Reserve maintained rates at 5.33\% through August 2024 before initiating a series of rate cuts beginning in September 2024, ultimately reducing the rate to 3.72\% by December 2025, a cumulative decline of 161 basis points. The token exhibited substantial primary market activity, with cumulative issuance of 10.22 million OUSG and redemptions totaling 5.07 million OUSG, resulting in net positive issuance of 5.15 million OUSG over the study period. Notable issuance spikes occurred in February 2025 (2.22M OUSG) and May 2025 (2.46M OUSG), potentially reflecting institutional portfolio rebalancing in response to the evolving interest rate environment.

\paragraph{Correlation with Monetary Policy}
We computed Pearson correlation coefficients between monthly changes in the Federal Funds Rate ($\Delta$FEDFUNDS) and on-chain trading volume (transfers excluding mint/burn transactions). The scatter plot in Figure~\ref{fig:govdebt} (bottom left) reveals a weak negative correlation ($r = -0.233$, $p = 0.284$, $n = 23$), suggesting that rate cuts are associated with modestly higher secondary market activity, though this relationship does not achieve statistical significance at the $\alpha = 0.05$ level.

The lack of strong correlation may be attributable to several factors: (1) OUSG is a restricted security available only to qualified purchasers, limiting secondary market liquidity; (2) the token's value derives from underlying Treasury yields rather than the Federal Funds Rate directly; and (3) institutional investors in tokenized treasuries may exhibit different trading patterns than retail cryptocurrency participants.

\paragraph{Net Capital Flows}
The net issuance flow panel (Figure~\ref{fig:govdebt}, bottom right) reveals alternating periods of capital inflows (positive net issuance) and outflows (redemptions exceeding new issuance). During the rate-cutting period (September 2024 onward, highlighted in blue), we observe continued net positive flows in aggregate, suggesting sustained institutional demand for tokenized Treasury exposure despite declining yields. This pattern is consistent with the hypothesis that blockchain-based Treasury products serve as an on-chain store of value and collateral instrument, with demand driven by factors beyond yield optimization alone.

\section{Related Work}

\noindent\textit{Asset Tokenization}. A growing body of literature examines asset tokenization as a mechanism for representing and transferring ownership or economic claims on distributed ledger infrastructures. Some studies conceptualize tokenization as the digital representation of existing assets or legally defined rights, emphasizing its potential to improve settlement efficiency, reduce intermediation costs, and expand market accessibility~\cite{wang2021sok,li2019tokenization,freni2020tokenization,gupta2020tokenization}. Other contributions focus on applying tokenization to broader asset categories, highlighting both operational benefits and regulatory challenges~\cite{cotler2023tokenized,avci2023blockchain,tian2020finance}.

\smallskip
\noindent\textit{Real-World Assets (RWAs)}. One prominent instantiation of these tokenization mechanisms is the RWA domain, which studies how off-chain assets and claims are brought into on-chain markets through compliant structuring and representation. A substantial body of work emphasizes the potential benefits of RWAs~\cite{saraswati2025real,aloshyna2025tokenization,lehmann2025utilizing}, framing tokenization as a bridge between traditional financial systems and blockchain-based markets. This literature highlights anticipated gains in capital efficiency, market accessibility, and composability, as well as the potential of RWAs to expand the scope of decentralized finance beyond crypto-native collateral. Related studies further examine the broader economic impact of RWA tokenization, noting that these effects depend on adoption, governance structures, and institutional integration \cite{baltais2024economic}.
Alongside these optimistic perspectives, another stream of research examines the systems-level implications of RWAs, focusing on the security, legal, market, and governance risks posed by linking on-chain tokens to off-chain assets~\cite{chen2024exploring,walmart2025integrating,xia2025exploration}. 
Lastly, a smaller but growing body of work focuses on the technical and architectural design of RWA systems, proposing protocol-level solutions to enhance interoperability, preserve privacy, and support cross-chain asset representation~\cite{guo2025xrwa,lee2025artex}.

\section{Conclusion and Prognostic Future Trends}
This paper has presented an overview of the emerging Real World Asset (RWA) concepts, analyzing the convergence of distributed ledger technology with traditional financial market structures. 
We have examined the tripartite framework defining this sector: the technical architectures required to bridge asynchronous systems, the financial mechanisms designed to integrate off-chain collateral with on-chain liquidity, and the evolving legal standards governing asset enforceability. 

\textit{Tokenization is Not Necessarily the Inevitable Shift:} 
The current industry obsession with tokenization may eventually prove to be a skeuomorphic end, a digital twin mirage where we simply replicate the inefficiencies of paper certificates onto a distributed database. 
Just as early internet interfaces mimicked physical desktops (skeuomorphism), if the current RWA protocols were merely files on a blockchain, inheriting all the legal friction of the underlying asset while adding new layers of technical complexity \cite{langley2024tokenization}.
The future, therefore, maybe not only about tokenizing legacy assets, but about the emergence of native digital assets that have no off-chain equivalent, i.e., assets whose economic substance derives from real-world activity or contractual relationships but whose creation, ownership, and settlement are defined entirely within on-chain systems rather than migrated from pre-existing legal or physical registries (not a Real-World Asset in the classical sense. But it can still represent real-world value or rights).
We cannot posit, but reasonably expect that the true endgame is not a world of thousands of fragmented RWA tokens floating on public chains, but the emergence of the unified ledger. 
As formalized by the Bank for International Settlements (BIS), this architecture does not move assets to a blockchain; rather, it moves the banking ledger onto a shared, programmable partition where central bank digital currencies (CBDCs) and commercial assets coexist \cite{bis2023blueprint}. 
In this singularity of finance, the concept of reconciliation becomes obsolete. 
There would be no non-atomic settlement, as asset transfer and payment would occur on a single, unified ledger rather than across two reconcilable databases; the transfer of the asset and the payment are executed as a single, atomic instruction code, rendering settlement risk a relic of the analog age \cite{carstens2023innovation}.
This shift will likely be driven not by crypto-native disruptors, but by the incumbents themselves co-opting the technology to reinforce, rather than replace, existing hierarchies. 
The recent move by SWIFT to integrate a shared ledger directly into its messaging stack suggests a future where the financial architecture does not migrate to a permissionless open sea of bearer tokens, but rather consolidates into a series of interconnected, highly regulated intranets \cite{swift2025ledger}. 
In this scenario, the tokenization narrative, which presumes the issuance of distinct digital assets on neutral public infrastructure, is superseded by a model of ledger development, where banks simply upgrade their internal databases to be programmable and interoperable without relinquishing control.
The future expectation, where invoices, royalty stream, and co-ownership of real estate are collateralized in real time, will likely occur within permissioned, trust-bounded liquidity domains, i.e., closed financial networks where asset access and interaction are restricted to compliant participants.
We argue that the trajectory suggests that tokenization may reinforce existing financial power structures, improving efficiency without necessarily advancing the decentralized principles (e.g., what Web3 community advocates for: permissionless access, self-custody, censorship resistance, and trust minimization) originally envisioned for blockchain systems.

\appendices

\newcolumntype{P}[1]{>{\RaggedRight\arraybackslash}p{#1}}

\begin{table*}[htbp!]
\centering
\caption{Comparison of Tokenized Real-World Asset (RWA) Projects}
\label{tab:rwa-projects}
\scriptsize
\begin{tabularx}{\textwidth}{
  P{0.08\textwidth}
  P{0.09\textwidth}
  P{0.09\textwidth}
  P{0.09\textwidth}
  P{0.09\textwidth}
  P{0.09\textwidth}
  P{0.08\textwidth}
  P{0.09\textwidth}
  P{0.14\textwidth}
}
\toprule
\textbf{Project} & \textbf{Asset Type} & \textbf{Backing} & \textbf{Governance} & \textbf{Blockchain(s)} & \textbf{Compliance} & \textbf{Audit/Attestation} & \textbf{Oracle / Data Source} & \textbf{Notes} \\
\midrule

Ondo (OUSG)\cite{ondo-ousg} & U.S. Treasuries & ETF / fund shares & Centralized & Ethereum & KYC (accredited/eligible) & Fund reporting & NAV updates & Tokenized money-market exposure. \\
BlackRock (BUIDL)\cite{blackrock-buidl} & U.S. T-bills/MMF & Fund shares (tokenized) & Centralized & Ethereum (+ multi-chain) & Qualified investors/KYC & Fund reporting & NAV/dividend feeds & Tokenized money market fund; on-chain share class. \\
Franklin Templeton (BENJI/FOBXX)\cite{franklin-benji} & U.S. govt MMF & Mutual-fund shares & Centralized & Stellar (+ others) & Regulated fund + KYC & Fund reporting & Daily NAV & On-chain recordkeeping for registered money market fund. \\
OpenEden (TBILL)\cite{openeden-tbill} & U.S. T-bills & Regulated fund shares & Centralized & Ethereum (EVM) & KYC/eligible investors & Regulated + reporting & Transparency/NAV reports & Tokenized T-bill fund with on-chain mint/redeem. \\
Circle/Hashnote (USYC)\cite{circle-usyc} & T-bills + repo & Fund shares (tokenized) & Centralized & Permissioned ERC-20 (EVM) & Whitelist/KYC-KYB & Fund reporting & NAV/price per share & Tokenized money market fund shares distributed on-chain. \\
Superstate (USTB)\cite{superstate-ustb} & Short-duration U.S. gov & Private fund shares & Centralized & Ethereum (EVM) & U.S. Qualified Purchasers + allowlist & Fund reporting & NAV feeds & Tokenized private fund exposure to gov securities. \\

Centrifuge\cite{centrifuge} & Private credit (SME) & Pool (DROP/TIN) & Hybrid (DAO + issuers) & Ethereum, Polkadot & Accredited/eligible & Audits vary & Issuer NAV feeds & Used as DeFi RWA collateral. \\
Goldfinch\cite{goldfinch} & Loans (emerging mkts) & Hybrid contracts & DAO (GFI) & Ethereum & KYC/KYB (varies) & Audits (smart contracts) & Borrower reports & Decentralized private credit. \\
Maple Finance\cite{maple} & Institutional credit & Partial collateral & DAO (MPL) & Ethereum, Avalanche & KYC for delegates & Audits (smart contracts) & Off-chain credit analysis & On-chain institutional lending. \\
TrueFi\cite{truefi} & Unsecured loans & Credit-based & Hybrid (TRU) & Ethereum & KYC for borrowers & Audits (smart contracts) & None / off-chain risk & Early unsecured DeFi credit. \\
Credix\cite{credix} & Private credit / receivables & Lending pools & Hybrid & Solana & KYB/KYC (borrowers/investors) & Audits (smart contracts) & Originator/servicer reports & Tokenized receivables/private credit pools. \\
Clearpool\cite{clearpool} & Institutional credit pools & Under-collateralized loans & DAO + pool operators & Ethereum, Polygon & Optional KYC (permissioned pools) & Audits (smart contracts) & Borrower disclosures & On-chain credit marketplace with pool delegates. \\

RealT\cite{realt} & Rental real estate & LLC equity tokens & Centralized & Ethereum, Gnosis & KYC/AML; Reg D/S & Internal/3rd-party varies & Off-chain valuation & Fractional rental income paid in stablecoins. \\
RedSwan CRE\cite{redswan} & Commercial real estate & SPV shares & Centralized (broker-dealer) & Ethereum, Hedera & SEC/FINRA (as applicable) & Custody/issuer audits & Off-chain NAV & Tokenized CRE offerings + secondary market. \\
Lofty AI\cite{lofty} & Residential real estate & Trust-backed tokens & Semi-centralized & Algorand & Reg CF + KYC & Audited (off-chain) & Off-chain valuation & Retail access; rent distribution. \\
Propy\cite{propy} & Real estate transactions & Deed/title workflows & Centralized platform & Ethereum (EVM) & Jurisdictional KYC/closing & Legal/closing docs & Property records & On-chain/NFT-style property transaction infrastructure. \\
tZERO / AspenCoin\cite{aspencoin} & Hotel/Resort equity & SPV equity token & Centralized & Ethereum (security token) & Securities compliance (Reg D etc.) & Offering/issuer reporting & Off-chain reporting & Tokenized St. Regis Aspen (security token case). \\
Tangible\cite{tangible} & Real estate (and RWAs) & NFT + custodial structures & Hybrid/DAO & Polygon (EVM) & KYC for redemption (varies) & Proof/reserve varies & Appraisal + off-chain feeds & Marketplace for tokenized tangible assets incl. RE. \\

Toucan (BCT)\cite{toucan} & Carbon credits & Retired credits & Semi-DAO & Polygon & Voluntary standards & Audited + traceability & Registry feeds & ReFi infra; credits bridged from registries. \\
Agrotoken\cite{agrotoken} & Commodities (grain) & Warehouse receipts & Centralized & Ethereum, Algorand & KYC for participants & Audited + IoT/ops & Index/oracle feeds & Tokenized grain receipts for LatAm supply chains. \\
Paxos (PAXG)\cite{paxos-paxg} & Gold & 1 oz LBMA gold & Centralized & Ethereum & Regulated issuer frameworks & Attestations (periodic) & Gold spot + reserve reports & Asset-backed gold token with published transparency reports. \\
Tether Gold (XAUt)\cite{tether-xaut} & Gold & 1 oz physical gold & Centralized & Ethereum (and others) & KYC for certain actions & Attestations (periodic) & Gold spot + issuer reports & Tokenized allocated gold with redemption program. \\
ComTech Gold (CGO)\cite{comtech-cgo} & Gold & 1g 999.9 gold & Centralized & XDC Network & KYC (platform) & Backing disclosures & Gold spot + issuer data & Tokenized gold instrument (often marketed as Shariah-compliant). \\
Meld Gold\cite{meld-gold} & Gold/Silver & Vault-backed tokens & Centralized & Algorand, XRPL & KYC (platform) & Audit/backing info & Metal spot + vault data & Tokenized precious metals with redemption rails. \\

Securitize\cite{securitize} & Securities & 1:1 digital securities & Centralized & Ethereum + others & Regulated TA / broker-dealer & Platform audits vary & None & Major tokenization platform (transfer agent rails). \\
Polymesh\cite{polymesh} & Regulated securities & Legal asset mapping & Hybrid (POLYX) & Polymesh L1 & On-chain identity/KYC & Multiple audits & Compliance embedded & Purpose-built L1 for regulated assets. \\
Swarm Markets\cite{swarm} & Stocks / ETFs & 1:1 custody & Centralized (MTF) & Ethereum, Polygon & BaFin-licensed & Custody + audit & AMM pricing & Regulated on-chain trading venue in EU context. \\
JPM Coin / Onyx\cite{jpm-coin} & Institutional deposits & 1:1 deposits & Centralized (bank) & Permissioned Ethereum/Quorum & Bank supervision & Internal controls & Internal data & Bank-led tokenized deposits/payments network. \\
HQLAx\cite{hqlax} & HQLA collateral swaps & Fully backed tokens & Centralized (consortium) & Corda R3 & Bank regulatory context & Legal/ops review & Custodian sync & DLT collateral mobility for high-quality liquid assets. \\

\bottomrule
\end{tabularx}
\end{table*}

\section{Representative RWA Protocol Implementations}
\label{sec:rwa-projects}
This subsection summarizes representative protocol designs (projects) within each major RWA asset class (see Table~\ref{tab:rwa-projects}), emphasizing how legal structuring, operational workflows, and on-chain interfaces jointly determine lifecycle execution.

\subsection{Tokenized Treasury and Fixed-Income Products}
Tokenized treasury and short-duration fixed-income products tend to converge on a fund or SPV wrapper model, where token holders own interests in a regulated vehicle that holds Treasury bills, repos, or related cash-equivalents. The implementation details then reduce to three recurring interfaces: (i) investor eligibility gating and transfer restrictions, (ii) valuation disclosure (typically NAV or price-per-share reporting), and (iii) redemption workflows that reconcile on-chain state with off-chain settlement rails. Ondo’s OUSG illustrates a treasury exposure product whose on-chain token represents an interest in an underlying fund/ETF-like structure and whose operational cadence is anchored in NAV-style reporting and institutional onboarding constraints \cite{ondo-ousg}. BlackRock’s BUIDL adopts the same economic primitive, a tokenized share class of a money-market style portfolio, but is optimized for institutional distribution, where corporate actions (e.g., dividends) and reporting are treated as first-order requirements rather than optional integrations \cite{blackrock-buidl}. Franklin Templeton’s on-chain government money fund (BENJI/FOBXX) is an important reference point because it shows how a registered fund administrator can treat blockchain as a transfer/record-keeping layer while retaining the conventional fund governance and disclosure stack \cite{franklin-benji}.

A second cluster in this category emphasizes streamlined mint/redeem while retaining a regulated wrapper. OpenEden’s TBILL product and Circle/Hashnote’s USYC both foreground the operational coupling between on-chain issuance limits and off-chain custody/reserve reporting, with token supply effectively bounded by the custodied portfolio state that must be attestable and auditable \cite{openeden-tbill,circle-usyc}. Superstate’s USTB similarly positions tokenized short-duration government exposure as a compliant, permissioned instrument whose value proposition is less "DeFi composability by default” than predictable settlement, reporting, and eligibility controls suitable for qualified purchasers \cite{superstate-ustb}. Across these designs, the technical differentiator is rarely the ERC-20 surface; it is the surrounding middleware: allowlists, NAV feeds, and operational controls that keep the on-chain representation isomorphic to the off-chain vehicle over time.

\subsection{Private Credit and Structured Financing}
Private credit and receivable-financing protocols introduce a different constraint: the underlying state is not a market price series but a servicing process. As a result, implementations are shaped by (i) how underwriting authority is assigned (DAO, delegates, or issuers), (ii) how defaults and recoveries are represented on-chain given that liquidation is legal and slow, and (iii) whether risk is redistributed through explicit tranching. Centrifuge is a canonical example of ``glass-box securitization'', where off-chain credit portfolios are financed through on-chain pools and split into senior/junior exposures (e.g., DROP/TIN) that embed a waterfall logic analogous to structured credit \cite{centrifuge}. Goldfinch and Maple both target institutional-style credit origination, but differ in how they operationalize risk selection: Goldfinch leans on borrower/operator reporting and protocol-level governance around credit lines, whereas Maple formalizes a delegate model that intermediates underwriting and monitoring under a protocol governance umbrella \cite{goldfinch,maple}. TrueFi represents an earlier and more aggressive design point that demonstrated under-collateralized (or unsecured) credit risk on-chain, thereby making counterparty assessment and loss management central to the protocol’s identity rather than ancillary features \cite{truefi}.

Two additional implementations illustrate the "pool marketplace” pattern. Credix emphasizes receivables and private credit pools with explicit originator and servicer dependencies, reflecting that the hard problem is not token minting but the integrity of off-chain cash-flow collection and reporting \cite{credix}. Clearpool structures borrowing through permissioned or semi-permissioned pools where borrower disclosures and pool operator governance serve as a practical substitute for on-chain liquidation guarantees, acknowledging that enforcement and recovery remain predominantly off-chain \cite{clearpool}. In aggregate, these protocols demonstrate that private-credit RWAs are best understood as synchronized socio-technical systems: smart contracts automate accounting and allocation, while underwriting, servicing, and enforcement remain anchored in institutions whose incentives must be made legible (and contestable) to on-chain participants.

\begin{table*}[htbp!]
\centering
\caption{Oracle Protocols in RWA Synchronization}
\label{tab:rwa_oracles}
\renewcommand{\arraystretch}{1.4}
\small
\begin{tabularx}{\textwidth}{@{}l l >{\raggedright\arraybackslash}X c@{}}
\toprule
\textbf{Protocol} & \textbf{Synchronization Model} & \textbf{Technical Mechanism \& Isomorphism Function} & \textbf{Ref.} \\
\midrule
\textbf{Chainlink} & Decentralized Push & \textit{Proof of Reserve (PoR).} Aggregates attests from multiple nodes querying custodian APIs to enforce cryptographic solvency (i.e., halting minting if $collateral < supply$). & \cite{breidenbach2021chainlink} \\
\midrule
\textbf{Pyth} & On-Demand Pull & \textit{First-Party Latency.} Publishes sub-second price updates directly from liquidity venues to the Wormhole message bus, enabling precise synthetic asset pricing. & \cite{pyth2024whitepaper} \\
\midrule
\textbf{API3} & First-Party Airnode & \textit{Serverless Bridging.} Allows institutional data providers (e.g., Open Banking APIs) to sign data at the source, eliminating third-party node operator risk. & \cite{api32020whitepaper} \\
\midrule
\textbf{UMA} & Optimistic Assertion & \textit{Dispute Arbitration.} Relies on economic guarantees (game-theoretic bonds) to resolve subjective states (e.g., insurance claim validity) via a challenge window mechanism. & \cite{uma2018whitepaper} \\
\midrule
\textbf{RedStone} & Modular Storage & \textit{Lazy-Loading.} Stores data signatures in Arweave and delivers them to the EVM only upon transaction execution, optimizing gas costs for long-tail assets. & \cite{redstone2023docs} \\
\bottomrule
\end{tabularx}
\end{table*}

\subsection{Real Estate and Income-Producing Assets}
Real-estate RWAs stress the operational boundary between local property management and global tradability. Most systems adopt an SPV/LLC equity-token model in which a legal entity owns the property and token holders own equity-like claims on net rental income and terminal proceeds. RealT is representative of the rental-income distribution model, where tokens map to fractional interests and recurring payouts are tied to the realized operating cash flow of the underlying properties rather than to a continuously marked-to-market price \cite{realt}. Lofty similarly targets retail accessibility and high-frequency distribution semantics (e.g., rent streaming), but the design remains constrained by jurisdictional compliance, property-manager workflows, and the realities of tenant turnover, maintenance, and expense reconciliation \cite{lofty}. RedSwan CRE illustrates the commercial-real-estate end of the spectrum, where broker-dealer style distribution, custody, and reporting are integral to the offering structure and where secondary liquidity is mediated through regulated venues and transfer restrictions \cite{redswan}.

A complementary set of projects focuses less on fractional rental yield and more on transaction infrastructure. Propy emphasizes deed/title workflows and closing rails, effectively treating tokenization as a mechanism for improving settlement and record synchronization in property transfer rather than as a pure yield instrument \cite{propy}. The AspenCoin case remains a useful reference for how hospitality or trophy assets can be wrapped in a security-token-like structure via an SPV, highlighting the limits imposed by securities law and the practical frictions of secondary market formation \cite{aspencoin}. Tangible represents a marketplace-oriented approach spanning multiple tangible RWAs (including real estate), combining custody and appraisal processes with on-chain representations that prioritize accessibility while necessarily inheriting the centralized trust assumptions of storage, redemption, and valuation \cite{tangible}. Together, these implementations underscore that real-estate RWAs are primarily about lifecycle engineering: property operations and legal enforceability dominate returns and risk, while on-chain logic mainly improves transparency, fractionalization, and settlement flexibility.

\subsection{Commodity-Backed Instruments}
Commodity-backed RWAs, particularly precious metals, compress the problem to custody and auditability: if the token is redeemable against vaulted bullion, then solvency becomes a question of whether reserves exist, are unencumbered, and are verifiably matched to supply. Paxos Gold (PAXG) and Tether Gold (XAUt) provide the most widely recognized patterns: token units correspond to specific amounts of physical gold held with custodians, while attestations and transparency reports serve as the bridge between off-chain reserves and on-chain liabilities \cite{paxos-paxg,tether-xaut}. ComTech Gold (CGO) and Meld Gold extend the same primitive to different custody networks and chains, illustrating that the main design degrees of freedom are not cryptographic but institutional: who the custodian is, how redemption is governed, and how audit evidence is produced and updated \cite{comtech-cgo,meld-gold}.

Beyond metals, commodity-like RWAs include environmental and resource rights where "backing” is a registry state rather than a warehouse inventory. Toucan’s carbon-credit tokenization demonstrates a registry-bridging model in which credits are represented on-chain with traceability anchored to off-chain registries and retirement (burn) events \cite{toucan}. Agrotoken provides an agricultural commodity reference point where backing is closer to warehouse receipts and supply-chain verification, and where the oracle problem becomes operational: the integrity of storage and receipt issuance determines the credibility of the on-chain claim \cite{agrotoken}. Across these instruments, the dominant risk is issuance--collateral mismatch; correspondingly, design emphasis shifts toward attestation frequency, custodian/registry credibility, and clear redemption or retirement semantics as the enforceable end state of the token.

\bibliographystyle{ieeetran}
\bibliography{reference.bib}

\end{document}